\documentclass[reprint,amsmath,amssymb, aps,]{revtex4-1}
\usepackage{graphicx}
\usepackage{dcolumn}
\usepackage{bm}
\begin{document}

\title{Coherent structure extraction in turbulent channel flow\\
using boundary adapted wavelets}
\thanks{This article is dedicated to Professor Javier Jimenez on the occasion of his 70th birthday.}

\author{Teluo Sakurai}
\affiliation{Department of Computational Science and Engineering, Nagoya University, Nagoya, 464-8603, Japan}
\author{Katsunori Yoshimatsu}
\affiliation{Institute of Materials and Systems for Sustainability, Nagoya University, Nagoya, 464-8603, Japan}
\author{Kai Schneider}
\affiliation{I2M-CNRS, Centre de Math\'ematiques et d'Informatique, \\ Aix-Marseille Universit\'e, 39 rue F. Joliot-Curie, 13453 Marseille Cedex 13, France}
\author{Marie Farge}
\affiliation{LMD--IPSL--CNRS, Ecole Normale Sup\'erieure, 24 rue Lhomond, \\75231 Paris Cedex 05, France}
\author{Koji Morishita}
\affiliation{Department of Computational Science, Kobe University, Kobe, 657-0013, Japan}
\author{Takashi Ishihara}
\affiliation{Center for Computational Science, Nagoya University, Nagoya, 464-8603, Japan}

\date{\today}

\begin{abstract}
We present a construction of isotropic boundary adapted wavelets, which are orthogonal and yield a multi-resolution analysis.
We analyze direct numerical simulation data of turbulent channel flow computed at a friction Reynolds number of 395,
and investigate the role of coherent vorticity.
Thresholding of the vorticity wavelet coefficients allows to split the flow into two parts, coherent and incoherent vorticity.
The coherent vorticity is reconstructed from their few intense wavelet coefficients.
The statistics of the coherent part, i.e., energy and enstrophy spectra, are close to the statistics of the total flow, 
and moreover, the nonlinear energy budgets are very well preserved.
The remaining incoherent part, represented by the large majority of the weak wavelet coefficients, corresponds to a structureless,
i.e., noise-like, background flow whose energy is equidistributed.
\end{abstract}


\maketitle

\section{Introduction}
Wall-bounded turbulent shear flows are of general interest in many engineering applications.
Three-dimensional (3D) turbulent channel flow bounded by two parallel walls is one of the canonical flow considered for direct numerical simulation (DNS).
Starting with the seminal work of Kim {\it et al}.~\cite{KMM},
many DNSs have been performed for increasingly higher Reynolds number, 
taking advantage of the growing power of supercomputers 
(see, e.g., a review article \cite{Smits}).
Currently the DNS with the highest friction based Reynolds number, $Re_\tau$, of $5200$ has been carried out by Lee and Moser~\cite{Lee}.

Turbulent flows are typically characterized by the excitation of a multitude of spatial and temporal scales, 
which involves a large number of degrees of freedom interacting nonlinearly.
Self-organization of the flow into coherent vortices is observed, 
even at large Reynolds number~\cite{BR},
where one observes that
these vortices are superimposed to a random background flow~\cite{She}. 
Moreover, turbulence exhibits significant spatial and temporal intermittency, 
especially in the dissipative range.
This implies that the strongest contributions become sparser and sparser while going to small scale in space and time. 
Wavelets being well-localized functions in space and in scale,
they yield efficient multi-scale decompositions, which have been applied to analyze, model and compute turbulent flows since 1988~\cite{FR,Mene91,Farge92,SV10}.
Decomposing turbulent flows into a wavelet basis yields a sparse representation,
namely the most energetic contributions are concentrated in few wavelet coefficients having strong intensity,
while the large majority of the remaining wavelet coefficients have negligibly small intensity.

The presence of coherent structures 
superimposed to a random background flow
motivated the development of the coherent vorticity extraction (CVE) method.
The idea of CVE, proposed by Farge {\it et al}.~\cite{FaPeSc2001,FaScKe99}, 
defines coherent structures as what remains after denoising the flow vorticity.
Vorticity is better localized in space than velocity, and thus more intermittent,
its wavelet decomposition is sparser and only few coefficients are necessary to represent the coherent structures.
The main reason is that, in contrast to the velocity, vorticity preserves Galilean invariance and has stronger topological properties owned to 
Helmholtz' and Kelvin's theorems.
Numerous applications of CVE can be found for periodic 
domains in the literature starting with homogeneous isotropic turbulence~\cite{FaScKe99,FaPeSc2001,FS01,FSPWR03pf,Vasi,OYSFK07}, 
temporally developing mixing layers~\cite{SFPR05jfm} and homogeneous shear flow with and without rotation~\cite{Jacobitz}.

For wall-bounded flows, 
the situation becomes more complex, 
because no-slip boundary conditions
have to be taken into account.
Indeed, no-slip boundary conditions generate vorticity due to the viscous flow interactions with the walls.
For turbulent boundary layers, Khujadze {\it et al}.~\cite{Khu} obtained an efficient algorithm to extract coherent vorticity, 
constructing a locally refined grid using wavelets with mirror boundary conditions.
However, this construction does not yield a multiresolution analysis where the basis functions are no more isotropic.
Fr\"{o}hlich \& Uhlmann ~\cite{FU} constructed wavelets based on second kind Chebyshev polynomials and applied them to channel flow data.
Scale-wise statistics in the wall-normal directions have thus been performed.
However, no fast wavelet transform (FWT) is available for these Chebyshev wavelet bases.
Two-dimensional (2D) wavelets have also been applied to wall-parallel planes in channel flows,
in order to examine turbulent statistics, 
in particular statistics of energy transfer~\cite{DM1,JR}.

The aim of the present work is to examine the role of coherent and incoherent flow contributions in 3D turbulent channel flow.
We propose a novel construction of 3D isotropic orthogonal wavelets using boundary wavelets in the wall-normal direction and periodic wavelets in the wall-parallel directions.
To this end, Cohen-Daubechies-Jawerth-Vial (CDJV) boundary wavelets~\cite{CDJV,CDV} having three vanishing moments,
and the periodized Coiflet 30 wavelets~\cite{Daub} having ten vanishing moments are employed.
These wavelets are orthogonal, 
the FWT can be used while taking into account boundary conditions, 
and the basis yields a multiresolution analysis.
Hence, the basis functions are isotropic since they have only one scale in all three spatial directions. 

DNS computation of the channel flow has been performed, and the data are analyzed at different time instants, 
using the above boundary adapted 3D isotropic wavelets.
The flow vorticity is decomposed into an orthogonal wavelet series, 
and we apply a thresholding to split the coefficients into two sets,
the coherent and incoherent ones.
The coherent vorticity, reconstructed from the few strongest wavelet coefficients, well preserves the turbulent statistics of the total flow, 
while the incoherent vorticity, reconstructed from the remaining large majority of the coefficients that are very weak, corresponds to a noise-like background flow.
The corresponding coherent and incoherent velocity fields are reconstructed from the coherent and incoherent vorticity fields, respectively, 
using the Biot-Savart relation satisfying the no-slip conditions at the walls.
Thus, we can efficiently examine the role of coherent vorticity in turbulent channel flow.
Other conventional methods, such as the $Q$-criterion and the $\lambda_2$ method~\cite{Jeong0,Jeong} 
could be used to identify coherent vortices in physical space,
as regions for which $Q$ or $\lambda_2$ is above a given threshold.
Here, $Q$ is the second-invariant of the 3D velocity gradient tensor, 
and $\lambda_2$ is the second largest eigenvalue of $S_{ij}S_{jk}+A_{ij}A_{jk} $,
where $S_{ij}$ and $A_{ij}$ are respectively the symmetric and antisymmetric tensor of the velocity gradient tensor.
It should be noticed that these quantities do not preserve the scale information about the vortices,
as the smoothness of the flow field is not preserved due to the clipping of vorticity in physical space. 
In contrast, the proposed wavelet filtering does preserve the smoothness of the coherent vorticity field and the multiscale properties of the coherent structures.

The paper is organized as follows:
Section \ref{sec2} presents the DNS computation and the data we analyze, including the methodology. 
The construction of isotropic wavelets is described, and the CVE method is summarized.
Numerical results are shown in Sec. \ref{sec3}.
Conclusions and perspectives are given in Sec. \ref{sec4}.

\section{\label{sec2}DNS and methodology}

\subsection{\label{sec2a}Direct Numerical Simulation}

\begin{figure}[tb]
\begin{center}
\includegraphics[width=80mm]{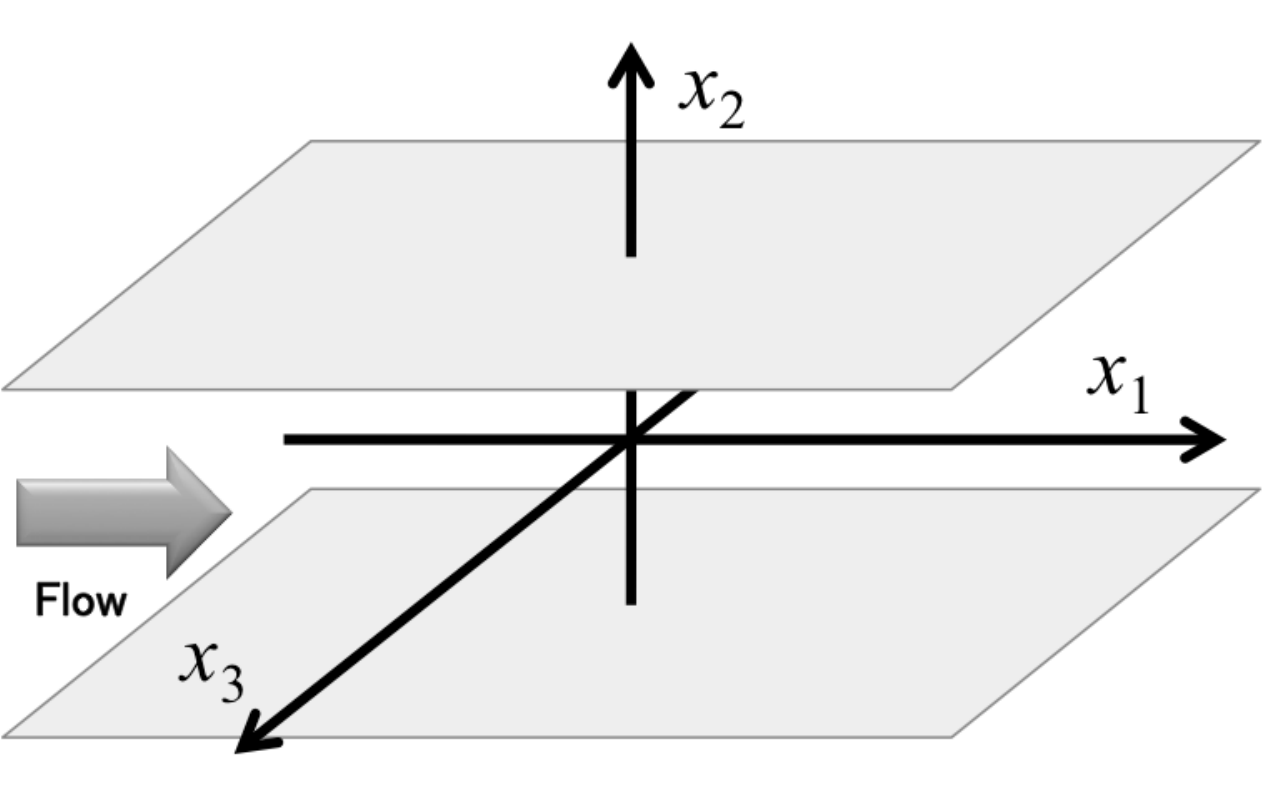}
\caption{Flow configuration for the turbulent channel flow.}
\label{coordinate}
\end{center}
\end{figure}

We consider 3D incompressible fluid flow in a channel bounded by two parallel walls subjected to a streamwise mean pressure gradient, 
which is a canonical flow configuration. 
It is illustrated in Fig. \ref{coordinate} together with the Cartesian coordinate system  $\bm{x}=(x_1,x_2,x_3)$, 
where the walls are at $x_2=\pm h$, $x_2$ being the wall-normal direction and $h$ the half width of the channel. 
The domain size in the streamwise $x_1$-direction is $2 \pi h$, 
and the size in the spanwise $x_3$-direction is $\pi h$. 
Periodic boundary conditions are respectively imposed in $x_1$- and $x_3$-directions,
while in the $x_2$-direction no-slip boundary conditions are satisfied at the walls.

The fluid flow motion obeys the Navier-Stokes equations with the incompressibility condition,
\begin{eqnarray}
& & \partial_t v_i + \partial_j (v_j v_i) = -\partial_i p + G \delta_{i 1}  + \nu \partial_j \partial_j v_i, \label{NSeq} \\
& & \partial_j v_j = 0, \label{divu}
\end{eqnarray}
where $v_i$ $(i=1,2,3)$ is the $i$-th velocity component,
$p$ is the pressure fluctuation, $G$ is the intensity of the mean pressure gradient in the $x_1$-direction, 
$\delta_{ij}$ is the Kronecker delta, 
$\nu$ is the kinematic viscosity, 
$t$ is time, and $\partial_t=\partial/\partial t$ and $\partial_i = \partial/\partial x_i$.
Einstein's summation convention is used for repeated indices.

We performed DNS of turbulent channel flow at $Re_\tau$ of $395$ using $N_1 N_2 N_3$ grid points, 
where $Re_\tau \equiv u_\tau h/\nu$, $u_\tau$ is the friction velocity defined by $[ \nu d U_1(x_2)/d x_2 ]^{1/2}$ at $x_2=-h$.
The velocity field $v_j$ is decomposed as $v_j=U_j(x_2) + u_j$ with $U_j$ being the mean velocity defined as 
$U_j= \langle v_i \rangle$, and $u_j$ are the velocity fluctuations. 
Here $\langle \cdot \rangle$ denotes the $x_2$-dependent spatial average of $ {\cdot} $ over the $x_1$-$x_3$ plane,
and $N_i$ is the number of the grid points in the $x_i$-direction, $N_1=N_3=256$ and $N_2=192$.
The toroidal and poloidal representation of Eqs.~(\ref{NSeq}) and (\ref{divu}) is employed, 
in order to satisfy the incompressibility constraint
as done by Kim {\it et al}.~\cite{KMM}. 
We used the Fourier pseudo-spectral method in the $x_1$-$x_3$ planes, 
and the Chebyshev-tau method in the $x_2$-direction.
The Chebyshev collocation points are given by $x_2=h \cos \{\pi(2 j+1)/(2 N_2)\}$ $(j=0, 1, \cdots, N_2-1)$ (see, e.g., Appendix B in Ref.~\cite{Canuto}). 
The aliasing errors are removed by the $3/2$ rule in the $x_1$-$x_3$ planes, 
and by the $2/3$ rule in the $x_2$-direction.
Time advancement is carried out using first-order implicit Euler method for the viscous terms, 
and a third-order Runge-Kutta method for the nonlinear terms and the mean pressure gradient term $G$ 
whose value is determined so that the total flow rate is kept constant.
The DNS code has been developed in Ref.~\cite{MorishitaETC13}.

Statistical quantities shown in this paper are obtained by time averaging over 40 DNS snapshots with intervals of $0.5$ washout time, defined by $ 2\pi h/U_1$.
The averaging starts after the total Reynolds stress, $- \langle u_1 u_2 \rangle + \nu d U_1/d x_2$ has become quasi-stationary.

\subsection{\label{sec2b}Wavelets}
\begin{figure*}[tb]
\begin{center}
\includegraphics[height=150mm]{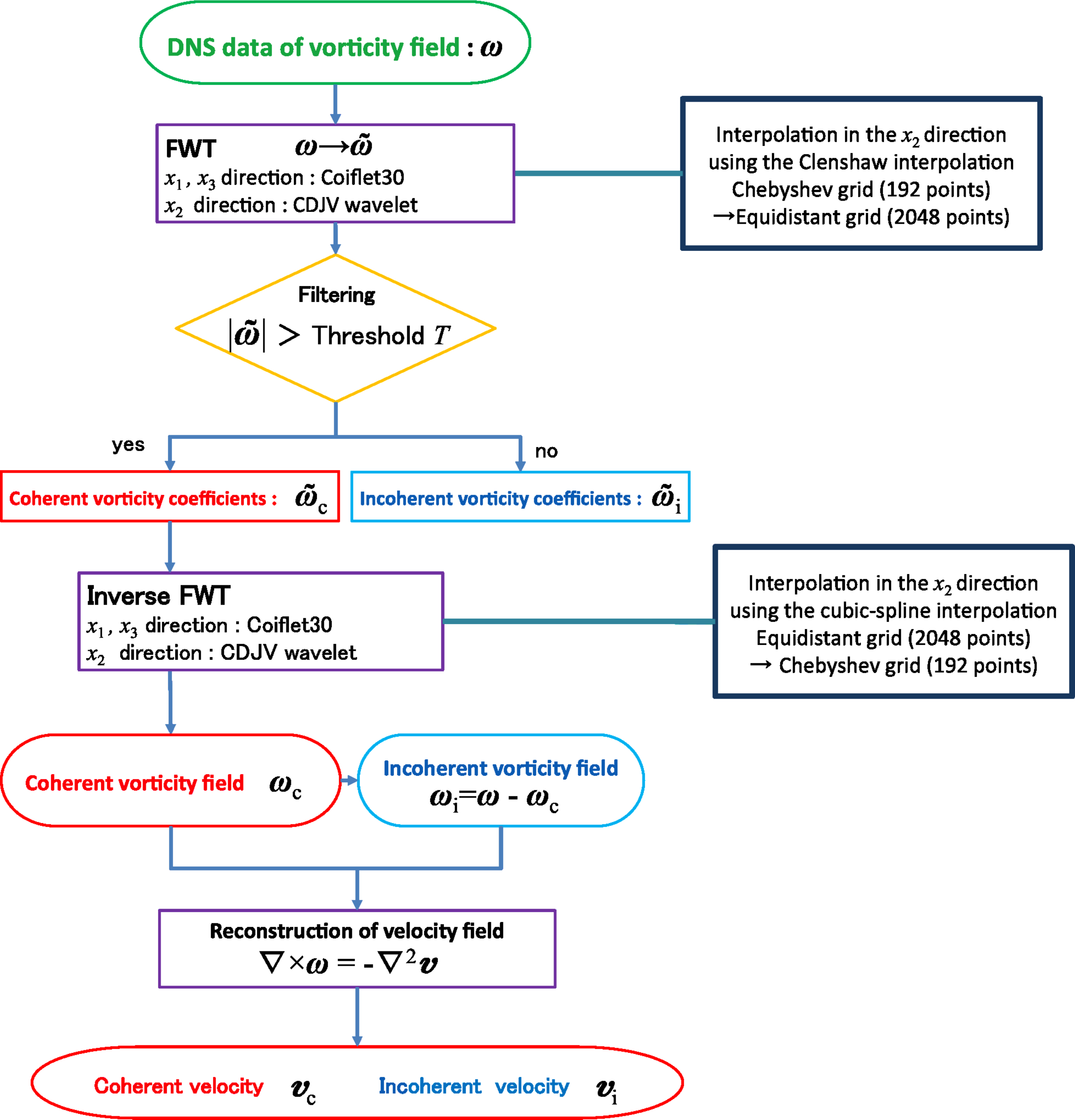}
\caption{Flowchart of the CVE procedure. }
\label{CVEalgor}
\end{center}
\end{figure*}

In this subsection,
we briefly summarize one-dimensional (1D) orthogonal periodized wavelets and 1D orthogonal boundary wavelets.
Then, we propose a 3D orthogonal isotropic wavelet transform constructed by tensor product of these 1D wavelets.
The CVE based on orthogonal wavelets to extract coherent vorticity out of turbulent channel flow is described in Sec. \ref{sec2c}.
In Fig. \ref{CVEalgor}, we present the flowchart of the CVE method used here.

\begin{figure*}[tb]
\includegraphics[width=80mm]{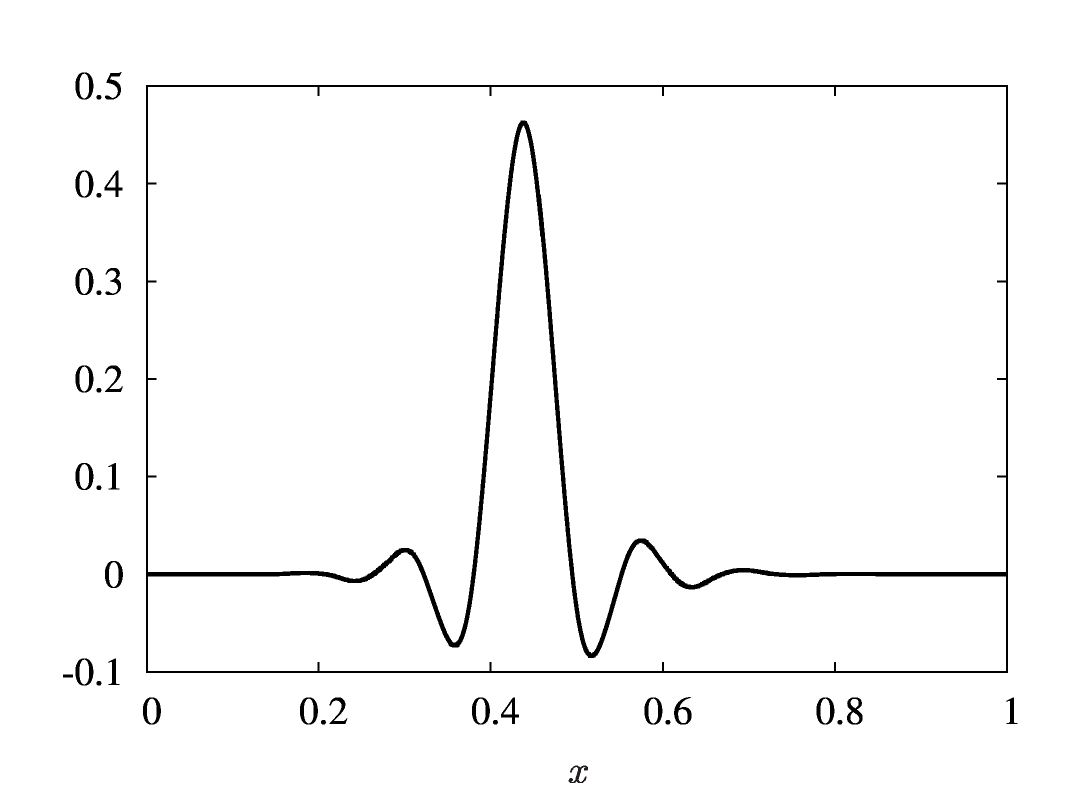}
\includegraphics[width=80mm]{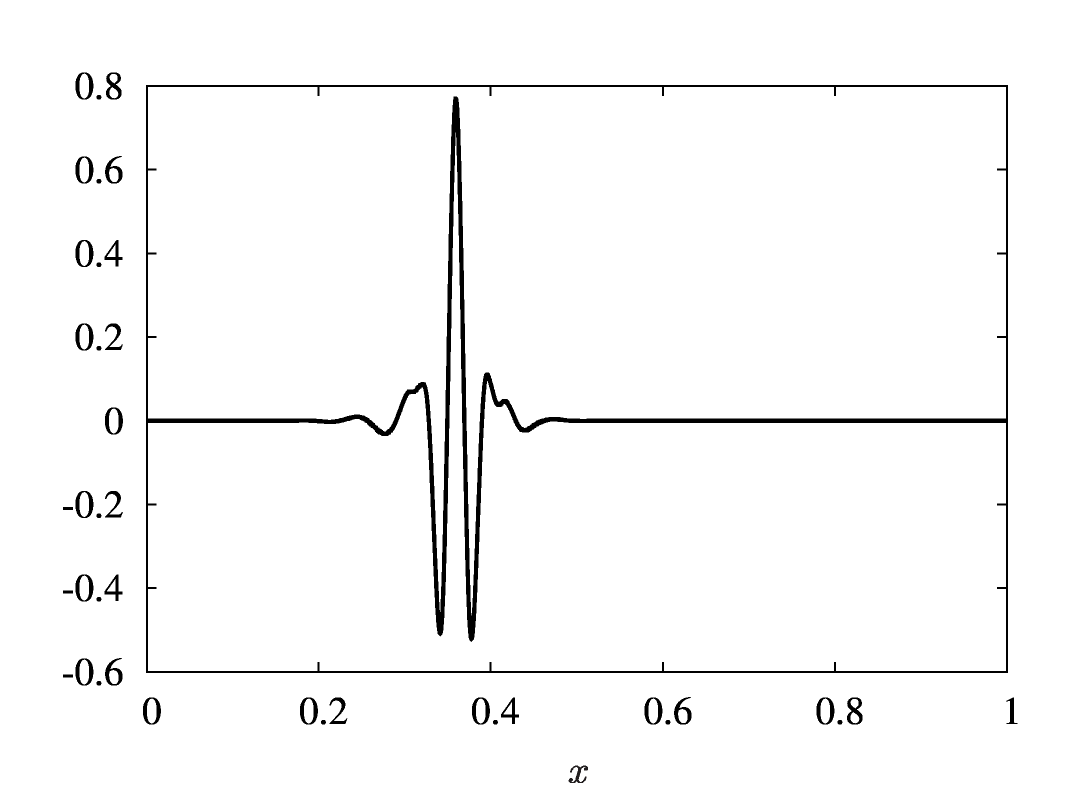}
\caption{Coiflet 30 wavelet on the periodic domain:  scaling function $\phi^P_{8,i}(x)$ (left) 
and corresponding wavelet $\psi^P_{8,i}(x)$ (right) both at scale $j=8$.}
\label{coiflet30}
\end{figure*}

\begin{figure*}[tb]
\includegraphics[width=80mm]{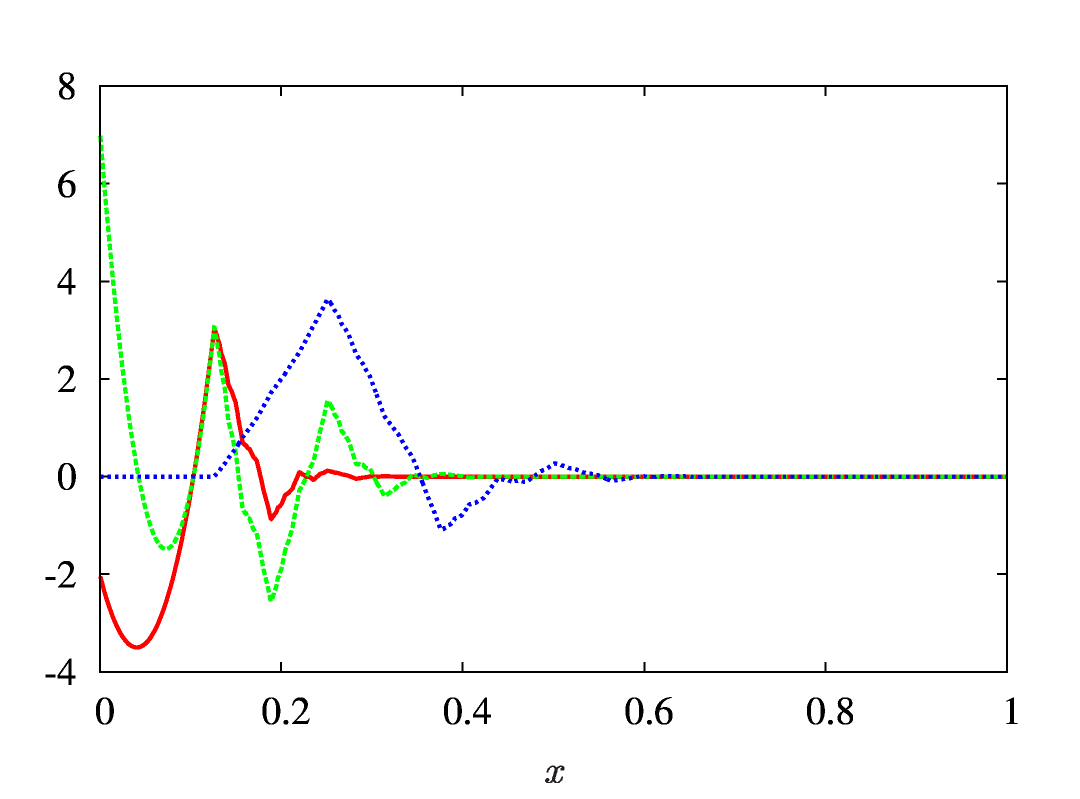}
\includegraphics[width=80mm]{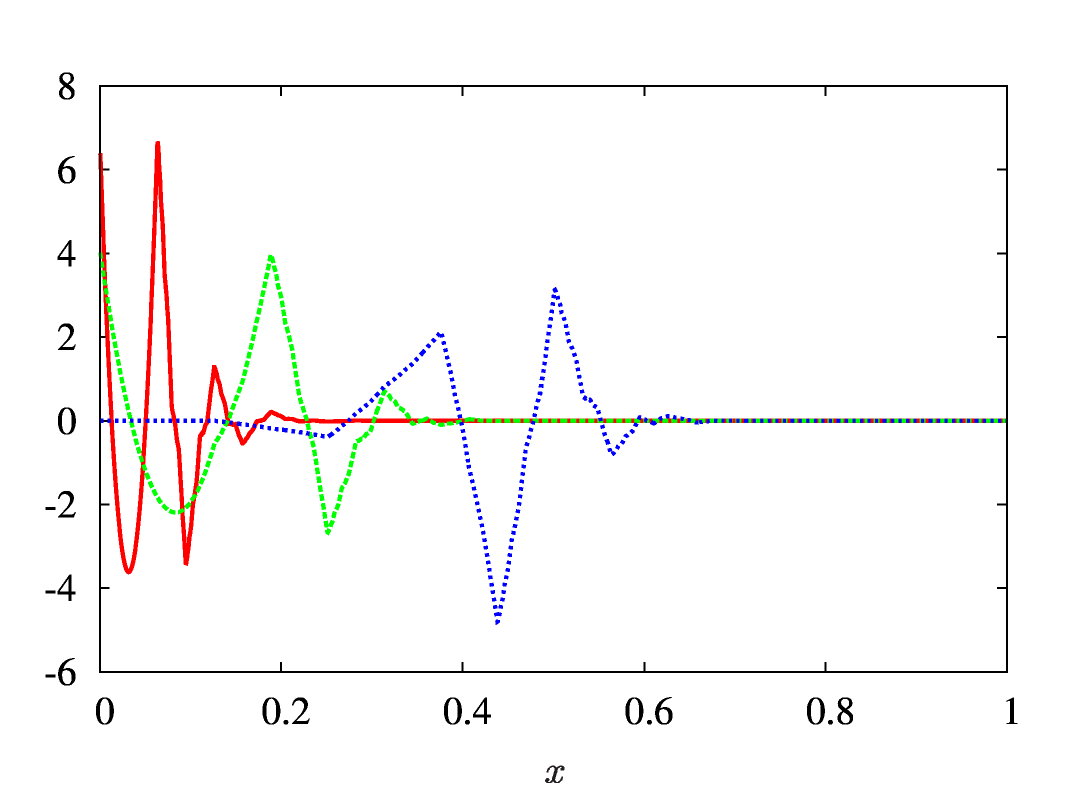}
\caption{CDJV wavelet on the interval: three scaling functions $\phi^B_{8,i}(x)$ (left) and three wavelets $\psi^B_{8,i}(x)$ (right) 
at scale $j=8$ and position $i=0$ (red), $63$ (green) and $127$ (blue) are shown. }
\label{cdjv}
\end{figure*}

We first consider 1-periodic wavelets $\psi^P(x)$ and their corresponding scaling function $\phi^P(x)$,
and orthogonal boundary adaptive wavelets $\psi^B(x)$ and their scaling function $\phi^B(x)$,
 with the boundaries at $x=(0,1)$.
Wavelets at scale $j$  are obtained by dilation, so that $\psi^\gamma_{j,0}(x)=2^{-j/2}\psi^\gamma(2^{-j}x)$ and 
$\phi^\gamma_{j,0}(x)=2^{-j/2}\phi^\gamma(2^{-j}x)$, where $\gamma=P, B$.
The periodized orthogonal wavelets are also self-similar with respect to translation.
Then the scaling function $\phi^\gamma$ and wavelet function $\psi^\gamma$ at scale $2^{-j}$ $(j\ge 0)$ and position $2^{-j}i$ $(i=0,1,\cdots,2^{j}-1)$, 
$\phi^P_{j,i}(x)$ and $\psi^P_{j,i}(x)$, 
are defined as $\phi^P_{j,i}(x)=2^{-j/2}\phi^P(2^{-j}x-2^{-j}i)$ and $\psi^P_{j,i}(x)=2^{-j/2}\psi^P(2^{-j}x - 2^{-j}i)$.
In contrast, $\psi^B_{j,0}$ and $\phi^B_{j,0}$ are no more translation invariant due to the boundary conditions, 
which modify the wavelets as position $i$ changes.
Readers interested in the details of boundary adapted wavelets may refer to the textbook~\cite{Mallat}, 
as the construction of $\psi^B_{j,i}$ and $\phi^B_{j,i}$ is rather technical.
All wavelets used here are orthonormal, i.e., 
$ \int_0^1 \psi_{j,i}^\gamma (x) \psi_{j',i'}^\gamma (x) d x =\delta_{ii'}\delta_{jj'} $,
$ \int_0^1 \phi_{j,i}^\gamma (x) \phi_{j,i'}^\gamma (x) d x =\delta_{ii'}$, 
and $ \int_0^1 \psi_{j,i}^\gamma (x) \phi_{j,i'}^\gamma (x) d x =0 $.

In this paper, we use Coiflet 30 wavelets~\cite{Daub} in the $x_1$- and $x_3$-directions, 
and the CDJV wavelets having 3 vanishing moments~\cite{CDJV,CDV} in the $x_2$-direction, 
both wavelets being compactly supported.
The Coiflet 30 wavelets are quasi-symmetric and have 10 vanishing moments.
The largest scale $2^{-j_0}$ of the CDJV wavelets satisfies $2^{j_0-1} \ge 3$~\cite{CDV}.
The illustrations of these wavelet functions are shown in Figs. \ref{coiflet30} and \ref{cdjv}.

The 3D orthogonal wavelets $\Psi^\mu$ $(\mu=1,2,\cdots,7)$ are obtained by tensor product such that
\begin{equation}
\begin{split}
& \Psi^\mu_{j,{\bm i}} (x_1,x_2,x_3) =\\
& \left\{
\begin{array}{lll}
\psi^P_{j,i_1}(x_1) \phi^B_{j',i_2}(x_2) \phi^P_{j,i_3}(x_3) & {\mathrm{for}} & \mu=1, \\
\phi^P_{j,i_1}(x_1) \psi^B_{j',i_2}(x_2) \phi^P_{j,i_3}(x_3) & {\mathrm{for}} & \mu=2, \\
\phi^P_{j,i_1}(x_1) \phi^B_{j',i_2}(x_2) \psi^P_{j,i_3}(x_3) & {\mathrm{for}} & \mu=3, \\
\psi^P_{j,i_1}(x_1) \phi^B_{j',i_2}(x_2) \psi^P_{j,i_3}(x_3) & {\mathrm{for}} & \mu=4, \\
\psi^P_{j,i_1}(x_1) \psi^B_{j',i_2}(x_2) \phi^P_{j,i_3}(x_3) & {\mathrm{for}} & \mu=5, \\
\phi^P_{j,i_1}(x_1) \psi^B_{j',i_2}(x_2) \psi^P_{j,i_3}(x_3) & {\mathrm{for}} & \mu=6, \\
\psi^P_{j,i_1}(x_1) \psi^B_{j',i_2}(x_2) \psi^P_{j,i_3}(x_3) & {\mathrm{for}} & \mu=7,
\end{array}
\right.
\end{split}
\end{equation}
where ${\bm i}=(i_1,i_2,i_3)$, $j'=j_0+j$ and $j=0,\cdots,J-1$.
The corresponding scaling function is defined as $\Phi(x_1,x_2,x_3)=\phi^P(x_1) \phi^B(x_2) \phi^P(x_3)$.

Now, let us consider a 3D vector field ${\bm w}({\bm x})=(w_1, w_2, w_3)$ in the computational domain $D$, 
where $D=\{x_1,x_2,x_3|0\le x_1 \le 2\pi h, -h \le x_2 \le h, 0\le x_1 \le \pi h \}$.
Before applying this wavelet decomposition, 
we interpolate ${\bm w}({\bm x})$ on an equidistant grid in the $x_2$-direction from the DNS data 
non-uniformly sampled on $N_2$ Chebyshev grid points in the wall-normal direction.
We thus get ${\bm w}({\bm x})$ uniformly sampled on $N_2'$ equidistant grid points at $x_{2,n}= h\{-1 + 2 n /(N_2'-1)\}$ $(n=0,\cdots,N_2'-1)$ 
using the Chebyshev interpolation~\cite{Clenshaw}. 
We choose $N_2'$ to be equal to $2048$ so that the flow field near the walls is kept well-resolved. 
We have $2/N_2' \sim  8\pi^2/N_2^2$, 
which shows that the grid width after the interpolation is comparable to the minimum grid width of the Chebyshev grid.
In the $x_1$- and $x_3$-directions,
we keep ${\bm w}({\bm x})$ uniformly sampled on $N_1$ and $N_3(=N_1)$ equidistant grid points, respectively.

The field ${\bm w}({\bm x})$, now sampled on $N_1 \times N_2'\times N_3 $ equidistant grid points, can then be decomposed into 
an isotropic orthogonal wavelet series as follows;
\begin{equation}
{\bm w}({\bm x}) = {\bar{\bm w}} + \sum_{\mu=1}^7 \sum_{j=0}^{J-1} \sum_{i_1,i_2,i_3=0}^{2^{j}-1} {\widetilde {\bm w}}^\mu_{j,{\bm i}} \Psi^\mu_{j,{\bm i}}({\bm x}) , 
\end{equation}
with wavelet coefficients computed with wavelets $\Psi^\mu_{j,{\bm i}}$
\begin{equation}
\begin{split}
& {\widetilde {\bm w}}^\mu_{j,{\bm i}} = \\
& \frac{1}{{\mathcal{V}}} \int_0^{L_1} d x_1 \int_{-h}^h d x_2  \int_0^{L_3} d x_3 \, 
{\bm w}({\bm x}) \Psi^\mu_{j,{\bm i}} \left( \frac{x_1}{2 \pi h}, \frac{x_2+h}{2h},\frac{x_3}{\pi h} \right), 
\end{split}
\end{equation}
\and the mean value computed with scaling function $ \Phi$
\begin{equation}
\begin{split}
& \bar{{\bm w}} = \\
& \frac{1}{{\mathcal{V}}} \int_0^{2\pi h} d x_1 \int_{-h}^h d x_2  \int_0^{\pi h} d x_3 \, {\bm w}({\bm x}) 
\Phi \left( \frac{x_1}{2 \pi h}, \frac{x_2+h}{2h} ,\frac{x_3}{\pi h} \right),
\end{split}
\end{equation}
where $J=\log_2 N_1$ and ${\mathcal{V}}=4 \pi^2 h^3$. 

\subsection{\label{sec2c}Coherent vorticity extraction}

\begin{figure}[tb]
\begin{center}
\includegraphics[width=80mm]{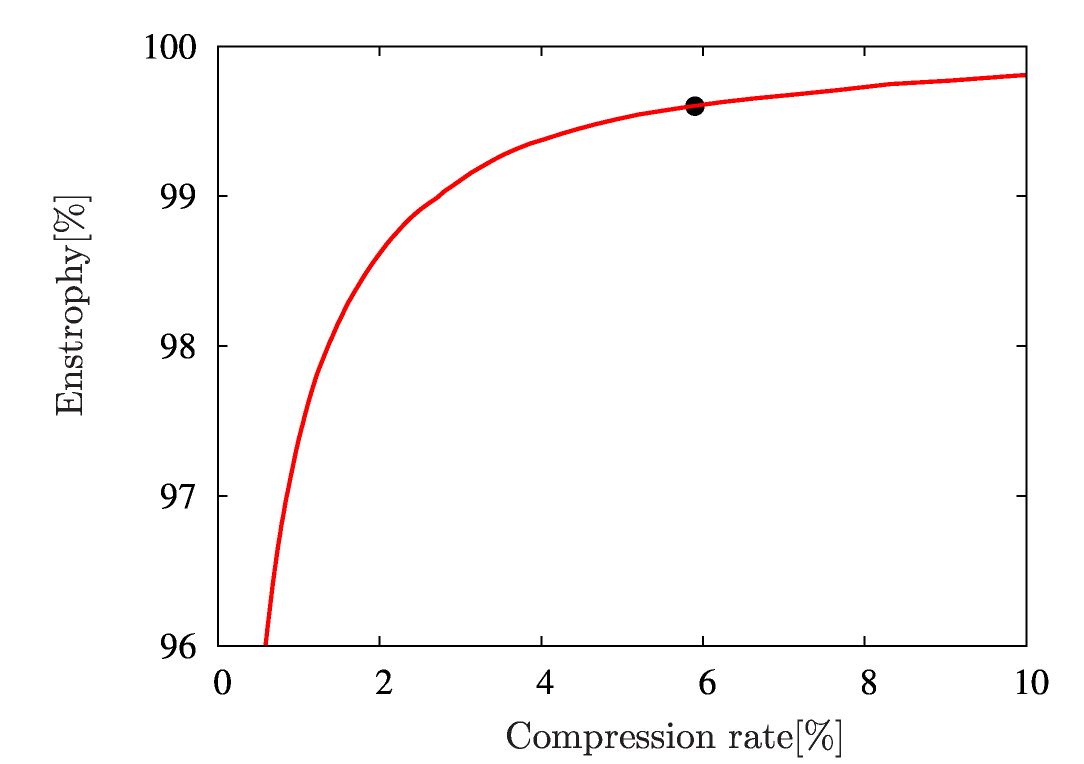}
\caption{Compression curve for the CVE: \% of retained enstrophy per unit volume vs. \% of retained wavelet coefficients. 
The circle corresponds to the threshold $T$ used for CVE in the following. }
\label{cr_vs_ens}
\end{center}
\end{figure}

We extract coherent vorticity out of turbulent channel flow data using the CVE method based on the wavelet decomposition of vorticity ${\bm \omega}(=\nabla \times {\bm v})$.
In the following, we summarize our method.
Then, since coherent structures do not have a universal definition yet, 
we define coherent structures as what remains after denoising. 
As first guess we consider the simplest type of noise, namely additive, Gaussian and white, i.e., uncorrelated.
Readers interested in details of this ansatz may refer to the original articles, e.g., Refs. ~\cite{FaPeSc2001,FaScKe99,OYSFK07}.

The CVE method is based on nonlinear thresholding the orthogonal wavelet coefficients of vorticity.
To this end the vorticity ${\bm \omega}$, interpolated on a sufficiently fine equidistant grid, is decomposed into an orthogonal wavelet series using the FWT.
Applying thresholding to the wavelet coefficients,
we split the flow into coherent and incoherent contributions.
The corresponding coherent and incoherent vorticity fields are then obtained by inverse wavelet transform.

In previous work, we used Donoho's threshold \cite{Donoho} to determine the value of the threshold and 
estimate the variance of the incoherent vorticity using an iterative scheme.
Azzalini {\it et al}. \cite{Azz} investigated the convergence of the iterative scheme 
and for isotropic turbulence Okamoto {\it et al.} \cite{OYSFK07} found that, 
depending on the Reynolds number, 8.7 \% and 6.0 \% are obtained for $Re_\lambda=167$ and $Re_\lambda=732$, respectively.
In Ref. \cite{FaPeSc2001}, Farge {\it et al.} used one iteration only, which was sufficient to get good compression while preserving the statistics of the total flow.
For the turbulent channel flow studied here we tried Donoho's threshold 
and found 
that very few wavelet coefficients keep almost the whole enstrophy of the flow, 
which is illustrated in the compression curve, shown in Fig. \ref{cr_vs_ens}. 
The flow visualization in Fig. \ref{vis_abs} 
shows tube-like coherent vortex structures of different intensity, which are very strong close to the wall and much weaker in the center of the channel. 
In the current work, we propose instead an {\it ad hoc} criterion for the threshold defined by 
$T=\langle |{\widetilde {\bm \omega}}^\mu_{j,{\bm i}}| \rangle_w +\alpha \langle \left(|{\widetilde {\bm \omega}}^\mu_{j,{\bm i}}| - \langle |{\widetilde {\bm \omega}}^\mu_{j,{\bm i}}| \rangle \right)^2 \rangle_w^{1/2}$,
where $\langle |{\widetilde {\bm \omega}}^\mu_{j,{\bm i}}| \rangle_w  = \sum_{\mu=1}^7 \sum_{j=0}^{J-1} \sum_{i_1,i_2,i_3=0}^{2^j-1} |{\widetilde {\bm \omega}}^\mu_{j,{\bm i}}|/(N_1N_2'N_3) $. 
Our aim is to retain only those wavelet coefficients which are responsible for the nonlinear dynamics of the flow, 
even if the fully developed turbulent regime has not been yet reached. 
We set $\alpha =0.75$ in the threshold value $T$ such that both velocity and vorticity statistics (as a function of $x_2$), 
together with the nonlinear dynamics and structures, are well preserved by the coherent flow. 

Using the inverse FWT, 
the coherent vorticity ${\bm \omega}_{c}$ is reconstructed from the wavelet coefficients whose intensity is larger than the threshold value $T$.
The incoherent vorticity ${\bm \omega}_{i}$ is then obtained using ${\bm \omega}_{i} = {\bm \omega} - {\bm \omega}_{c}$.
To get ${\bm \omega}_{c}$ and ${\bm \omega}_{i}$ sampled on the Chebyshev grid points, which are useful and efficient for data analysis presented in Sec. \ref{sec3}, 
we perform a cubic spline interpolation in the $x_2$-direction.

Owing to the orthogonality of the wavelet decomposition, ${\bm \omega}_c $ is orthogonal to ${\bm \omega}_i $ and thus $Z_t=Z_c+Z_i$, 
where $Z_t$, $Z_c$ and $Z_i$ are respectively the total, coherent and incoherent enstrophy per unit volume,
defined as $Z_\alpha =\int \!\! \int \!\! \int_D |{\bm \omega}_\alpha|^2  d {\bm x}  /(2 {\mathcal{V}} )$ ($\alpha=t, c, i$).
The coherent velocity ${\bm v}_c$ and the incoherent velocity ${\bm v}_i$ are computed from ${\bm \omega}_c$ and ${\bm \omega}_i$ by solving Biot-Savart's relation, $\nabla^2 {\bm v}=-\nabla \times {\bm \omega}$, respectively.
It is noted that ${\bm v}_i$ and ${\bm v}_c $ are weakly non orthogonal, i.e., the cross term  $ \int {\bm v}_i \, {\bm v}_c \, d{\bm x}$ is below $0.4 \%$ of the total energy.

\section{\label{sec3}Numerical results}

Now we analyze 40 snapshots of DNS data for the turbulent channel flow with intervals of 0.5 washout times,
and we ensemble-average over those 40 snapshots to guarantee well-converged statistical results.
We examine contributions of coherent and incoherent flows obtained with the previously described CVE method.
Quantities with the superscript $^+$ are expressed in wall units, i.e., 
they are non-dimensionalized by $u_\tau$ and $\nu$.
We define the distance from the wall $y$ as $y=x_2+1$. 

\subsection{Visualization}

\begin{figure*}[tb]
\begin{tabular}{c|c}
\begin{minipage}{0.5\textwidth}
\begin{center}
\includegraphics[width=70mm]{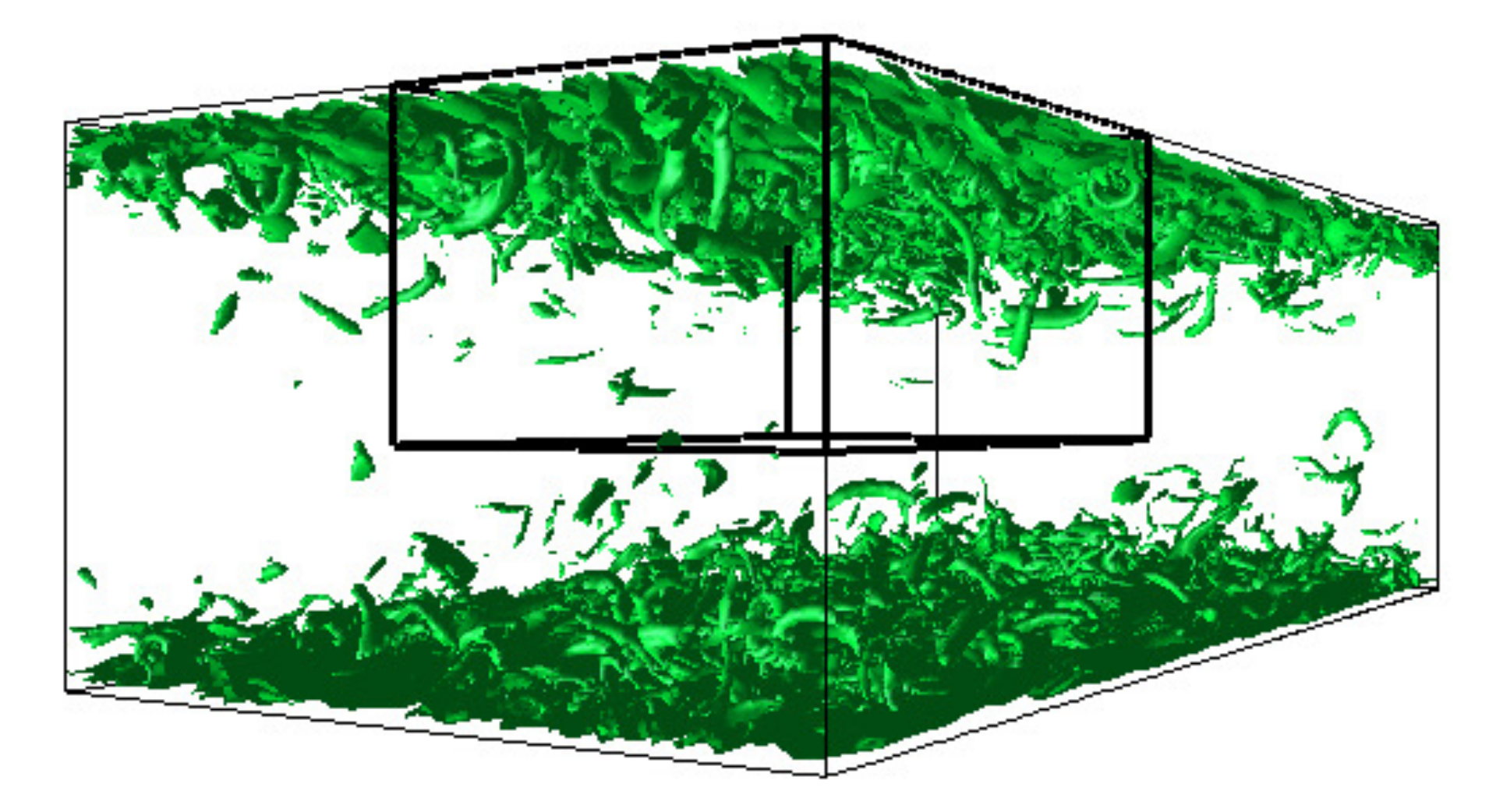}\\
\includegraphics[width=70mm]{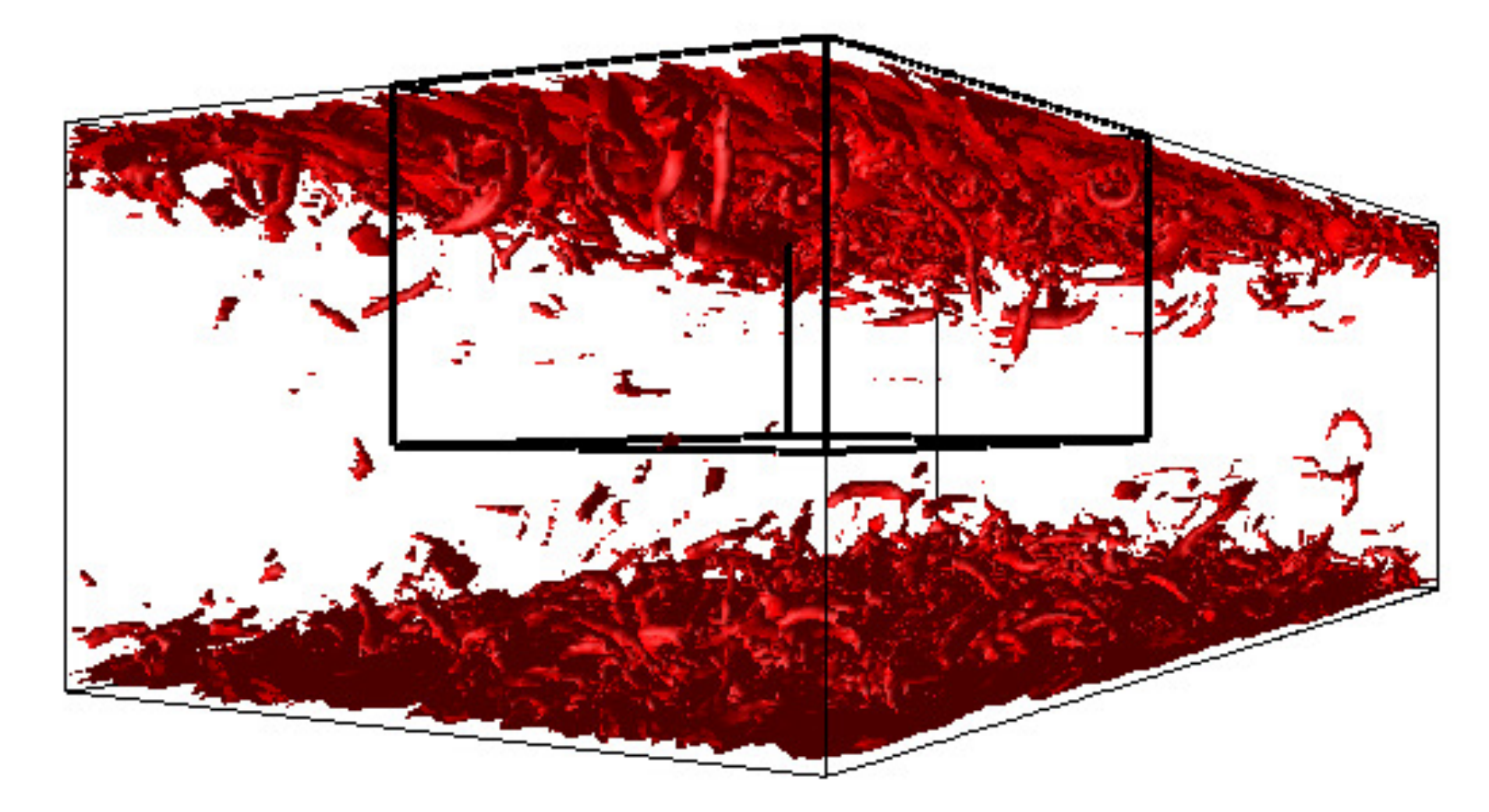}\\
\includegraphics[width=70mm]{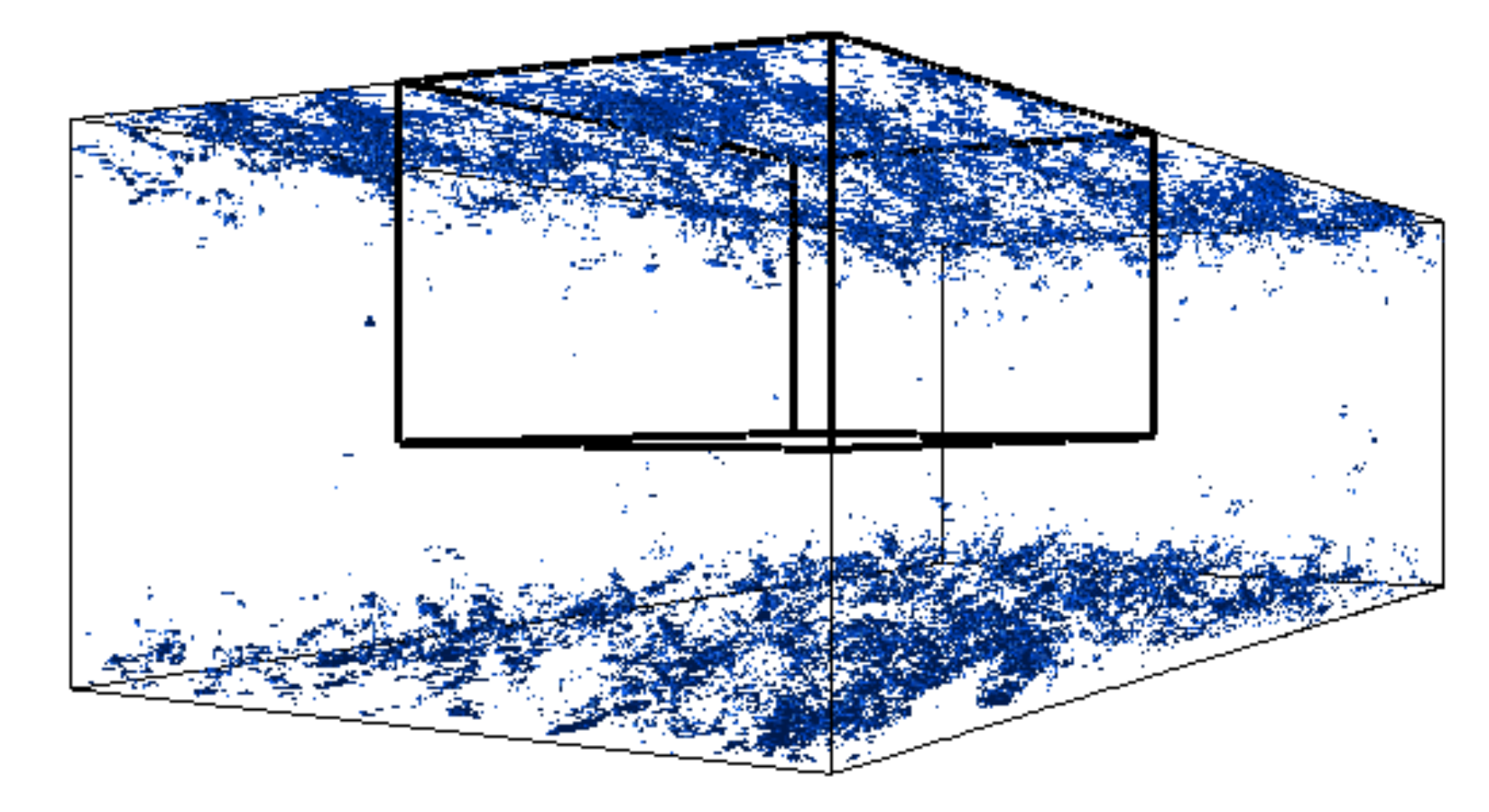}
\end{center}
\end{minipage}
&
\begin{minipage}{0.5\textwidth}
\begin{center}
\includegraphics[width=60mm]{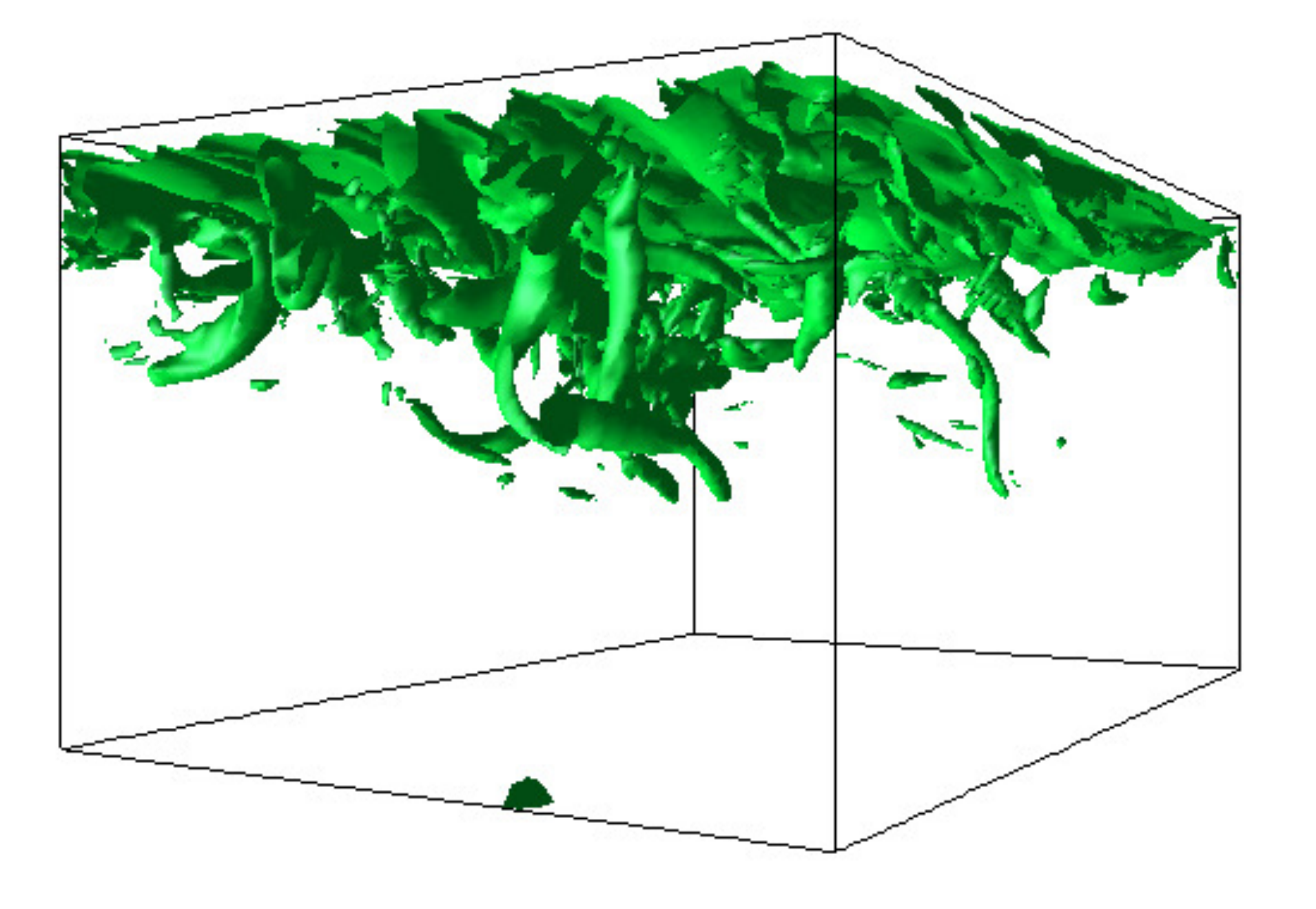}\\
\includegraphics[width=60mm]{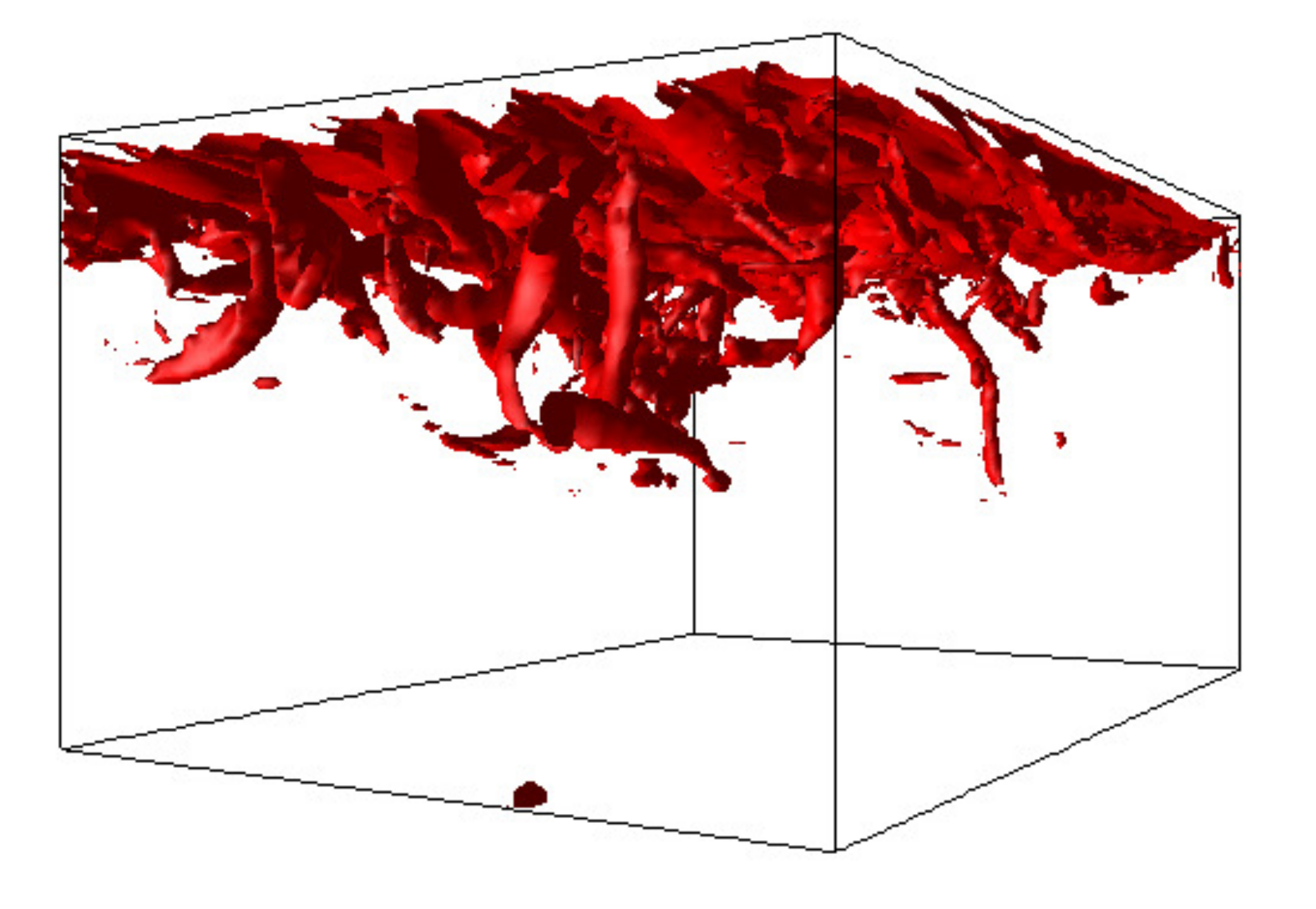}\\
\includegraphics[width=60mm]{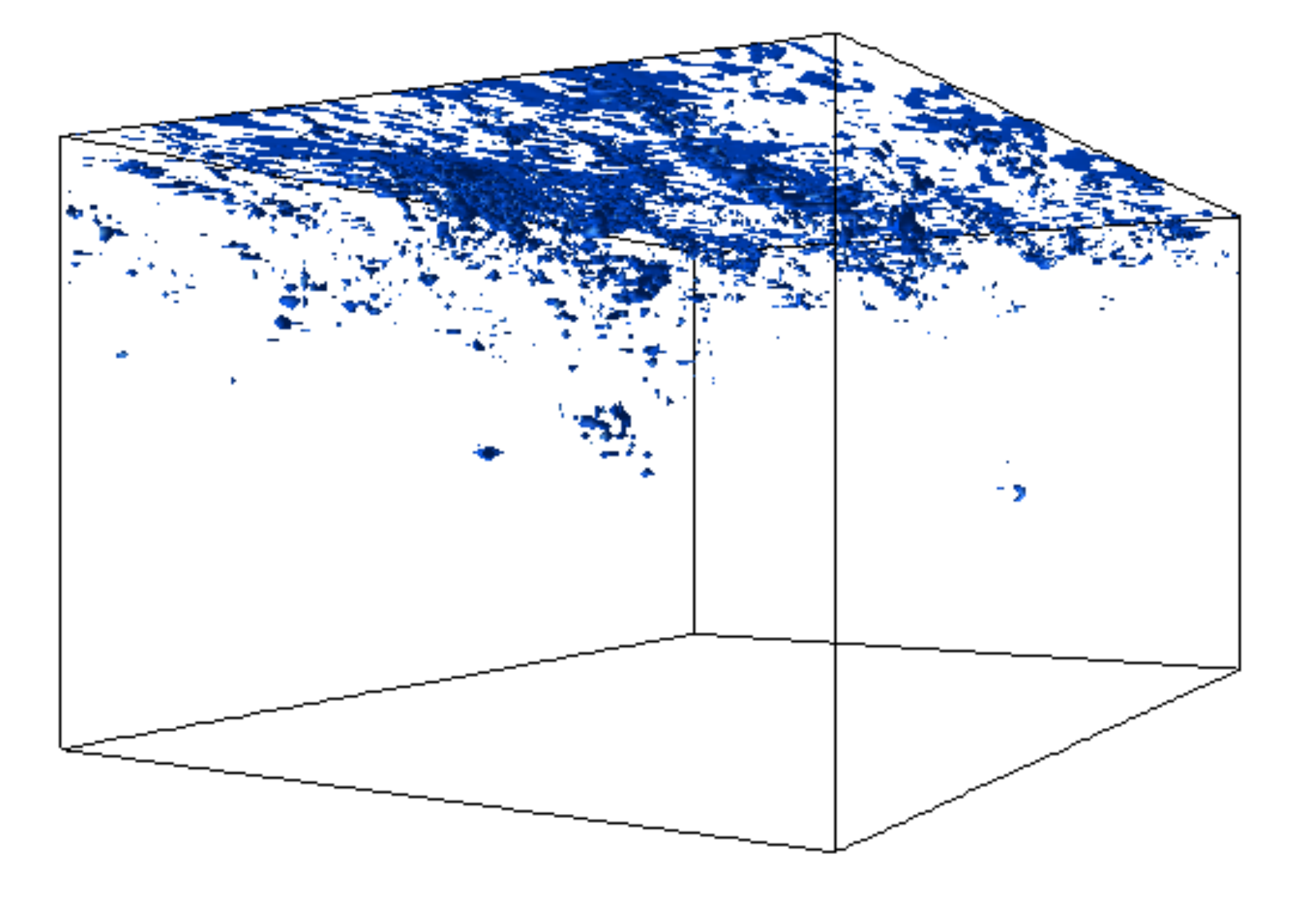}
\end{center}
\end{minipage}
\end{tabular}
\caption{Visualization of total vorticity $\bm{\omega}$ (green), coherent vorticity $\bm{\omega}_c$ (red) and incoherent vorticity $\bm{\omega}_i$ (blue).
The left column presents isosurfaces $|\bm{\omega}^{+}|=|\bm{\omega}_c^{+}|=0.3$ and $|\bm{\omega}_i^{+}|=0.1$.
The right column shows their zooms in the wall region where $0 \le x_1 \le 0.71\pi h$, $ -0.12 h \le x_2 \le h$, and $0.5\pi h \le x_3 \le \pi h$.}
\label{vis_abs}
\end{figure*}

\begin{figure*}[tb]
\begin{tabular}{c|c}
\begin{minipage}{0.5\textwidth}
\begin{center}
\includegraphics[width=70mm]{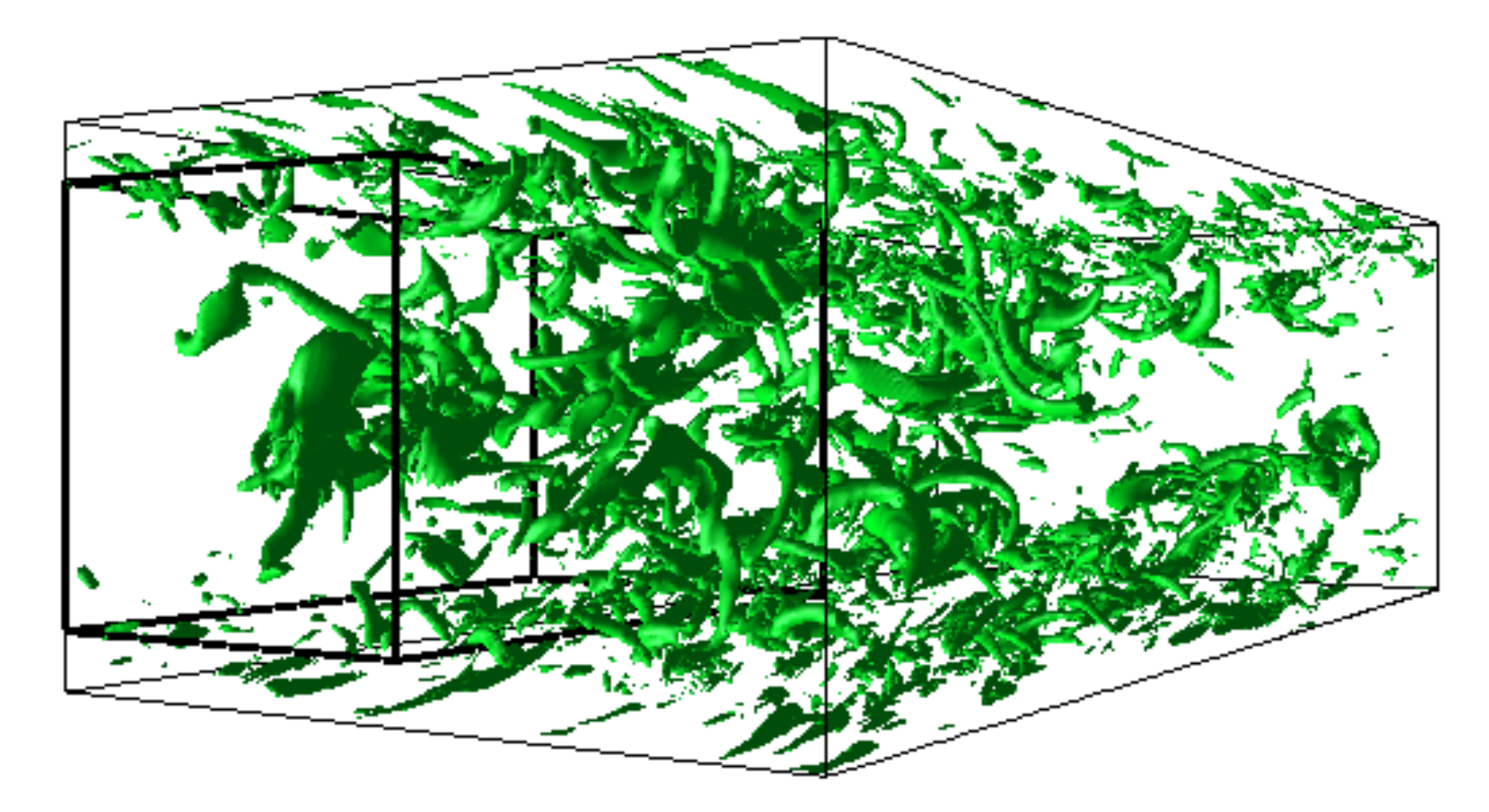}\\
\includegraphics[width=70mm]{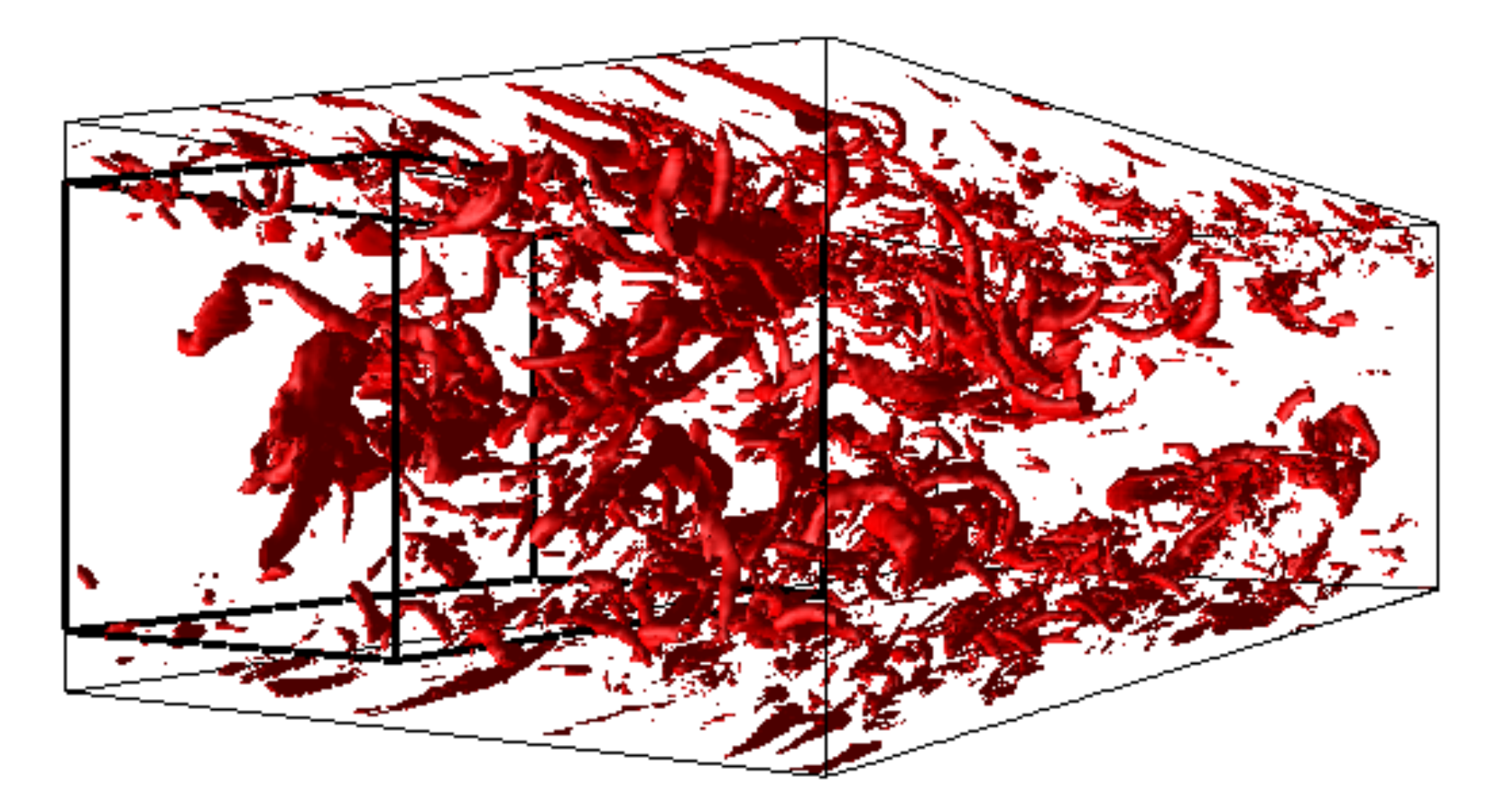}\\
\includegraphics[width=70mm]{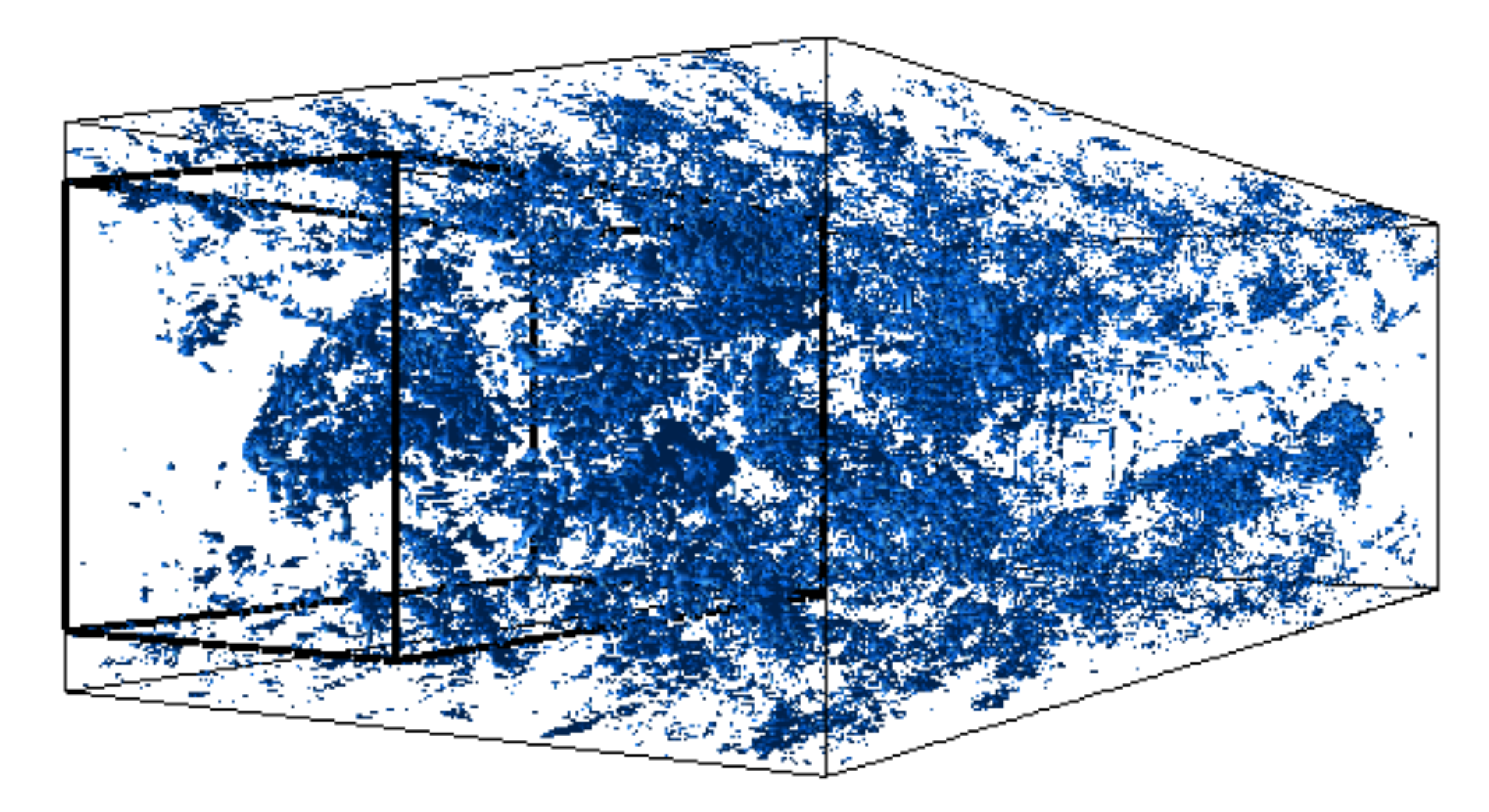}
\end{center}
\end{minipage}
&
\begin{minipage}{0.5\textwidth}
\begin{center}
\includegraphics[width=45mm]{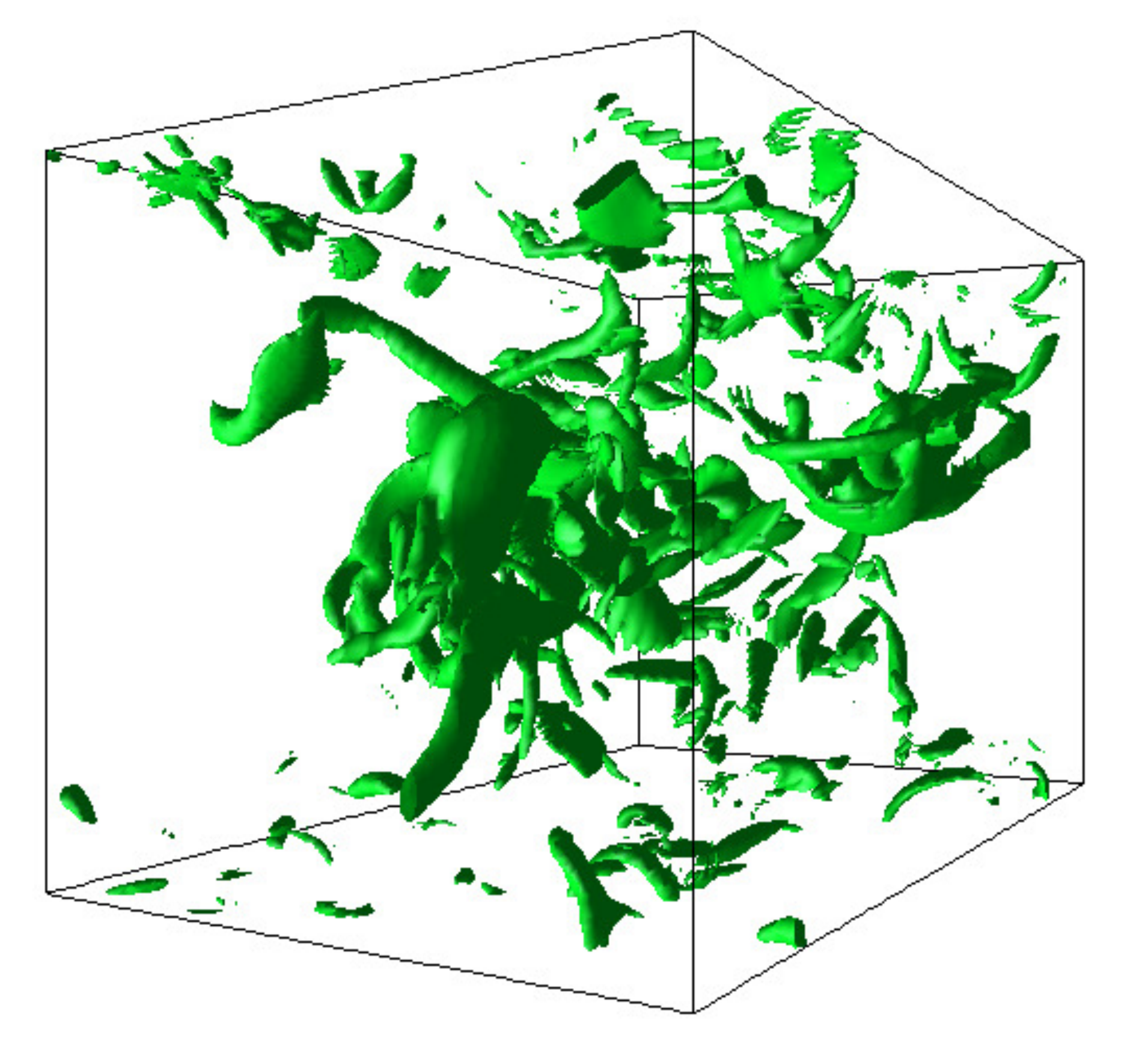}\\
\includegraphics[width=45mm]{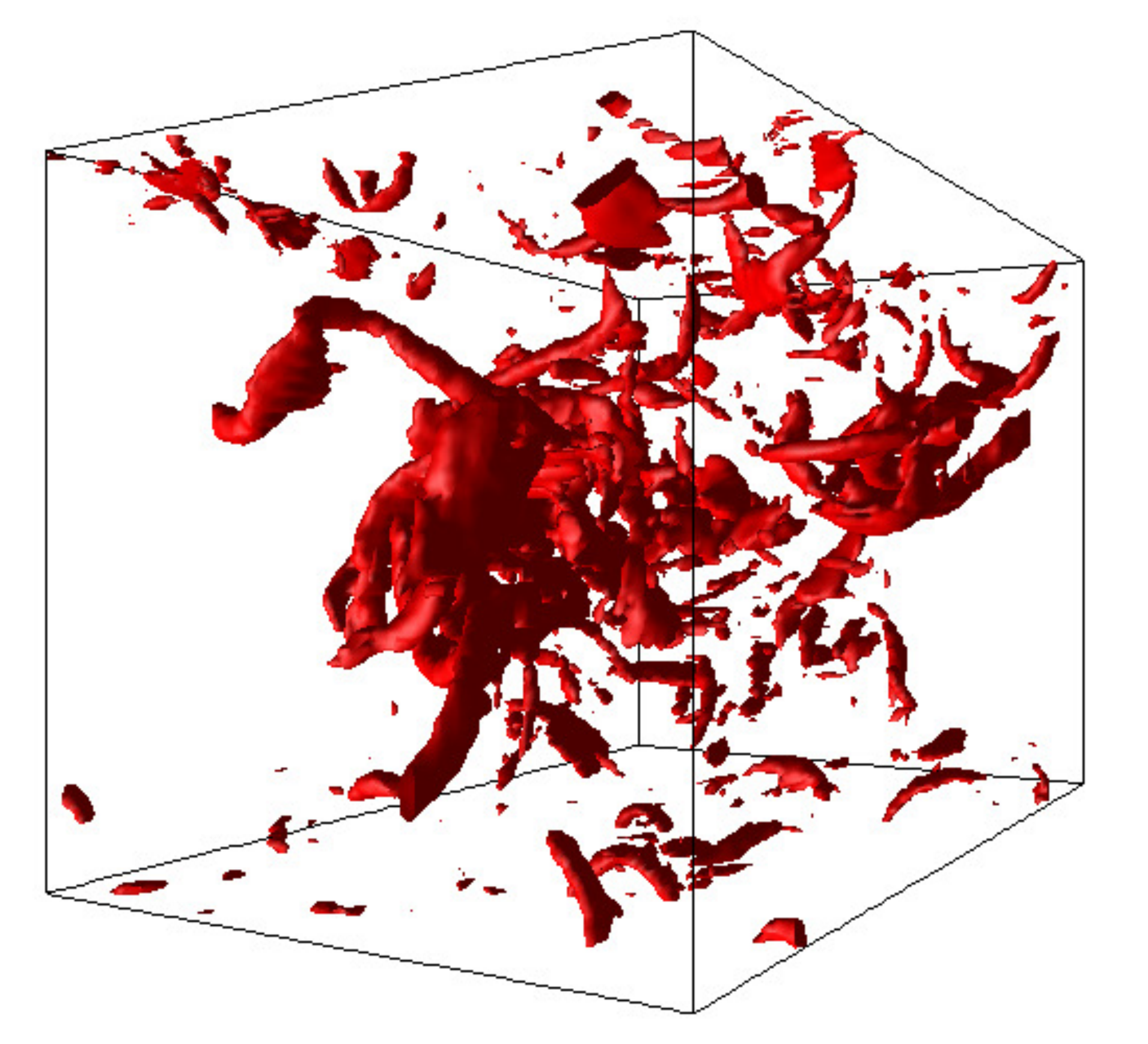}\\
\includegraphics[width=45mm]{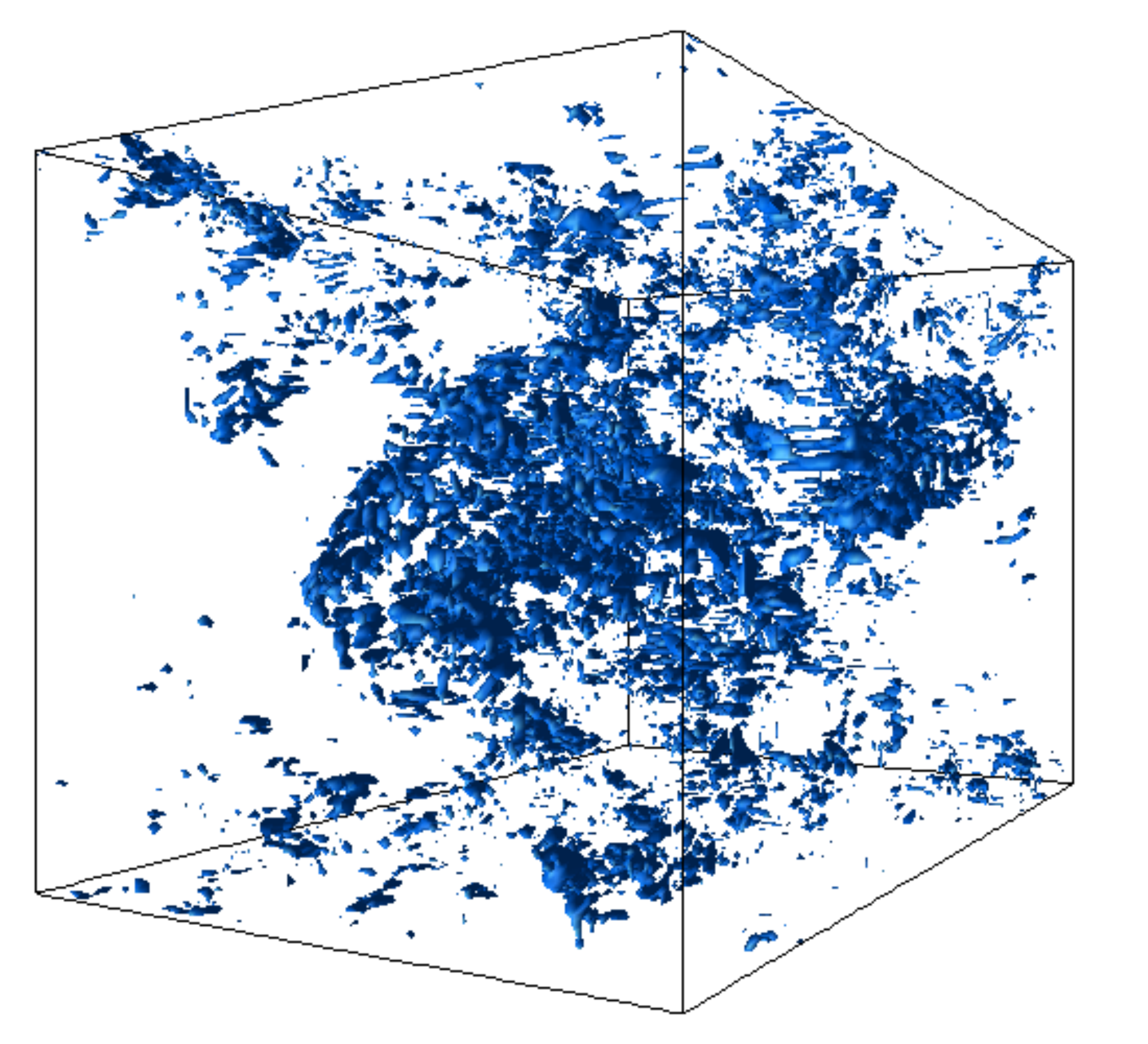}
\end{center}
\end{minipage}
\end{tabular}
\caption{Visualization of total vorticity $\bm{\omega}$ (green), coherent vorticity $\bm{\omega}_c$ (red) and incoherent vorticity $\bm{\omega}_i$ (blue).
Isosurfaces $|\bm{\omega}^{+}|=|\bm{\omega}_c^{+}|=\langle|\bm{\omega}^{+}|\rangle+3\langle(|\bm{\omega}^{+}|-\langle|\bm{\omega}^{+}|\rangle)^2\rangle^{1/2}$ and 
$|\bm{\omega}_i^{+}|=\langle|\bm{\omega}_i^{+}|\rangle+3\langle(|\bm{\omega}_i^{+}|-\langle|\bm{\omega}_i^{+}|\rangle)^2\rangle^{1/2}$ are shown.
The right column presents corresponding zooms in the core where $ 0 \le x_1 \le 0.79\pi h$, $-0.78 h \le x_2 \le 0.78 h$ and $ 0 \le x_3 \le 0.5\pi h$.}
\label{vis_sd}
\end{figure*}

Visualizations of isosurface values of the modulus of vorticity for the total, coherent and incoherent flows given 
at the same time instant are shown in Figs. \ref{vis_abs} and \ref{vis_sd}.
Corresponding zooms are also presented to see flow structures more clearly.
Figure \ref{vis_abs} shows that the most intense vorticity structures are near the walls.
Since the incoherent vorticity is much weaker than the total and coherent vorticities in Fig. \ref{vis_abs},
the isosurface value for the incoherent vorticity ${\bm \omega}_i$ is reduced by a factor 3 compared to the coherent and total vorticities.
On the other hand, Fig. \ref{vis_sd} visualizes vorticity structures in the core region, using $y^+$-dependent isosurface values,
$|\bm{\omega}^{+}|=|\bm{\omega}_c^{+}|=\langle|\bm{\omega}^{+}|\rangle+3\langle(|\bm{\omega}^{+}|-\langle|\bm{\omega}^{+}|\rangle)^2\rangle^{1/2}$ 
and  $|\bm{\omega}_i^{+}|=\langle|\bm{\omega}_i^{+}|\rangle+3\langle(|\bm{\omega}_i^{+}|-\langle|\bm{\omega}_i^{+}|\rangle)^2\rangle^{1/2}$, 
recalling $\langle \cdot \rangle$ denotes the $y^+$-dependent spatial average of $ {\cdot} $ over each wall-parallel plane.

We observe that the total flow exhibits intense vortex tubes near the walls,
as in previous DNS (e.g., Ref. ~\cite{Blackburn}),
but we also see them in the core region, however they are less intense.
Looking at the coherent flow, we find that these tubes are well preserved by ${\bm \omega}_c$,
which is reconstructed from only $ 0.55\% $ of the $256^2 \times 2048(\simeq 13 \times 10^7)$ wavelet coefficients, 
i.e., $5.9\%$ of the original $256^2 \times 192(\simeq 1.2 \times 10^7)$ grid points.
The coherent flow retains almost all of the total energy and enstrophy, 
$99.9 \%$ of the total energy and $99.7 \%$ of the total enstrophy.
In contrast, the incoherent vorticity ${\bm \omega}_i$ looks less organized without exhibiting vortex tubes near the walls and in the core region.
Although the incoherent flow is represented by the remaining majority of wavelet coefficients,
it retains a negligible amount of energy, namely $2.3\times10^{-3} \%$ of the total energy, and only $0.5 \%$ of total enstrophy. 

\subsection{Mean velocity and vorticity statistics}

\begin{figure*}[tb]
\includegraphics[width=80mm]{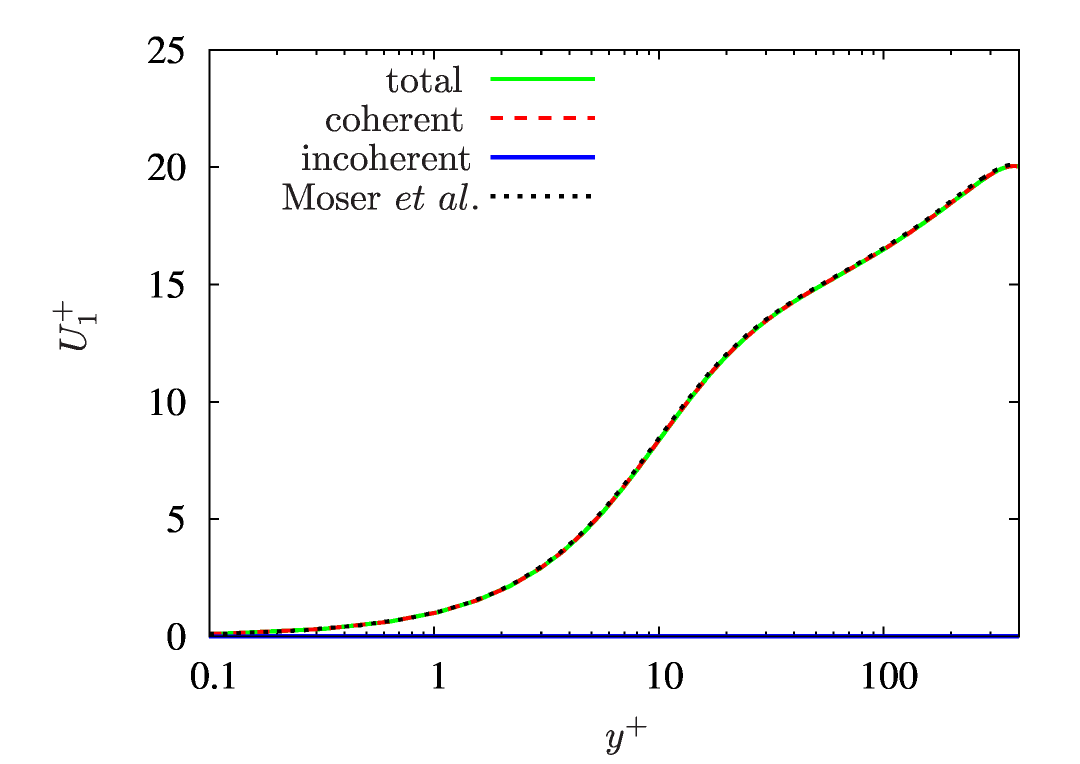}
\includegraphics[width=80mm]{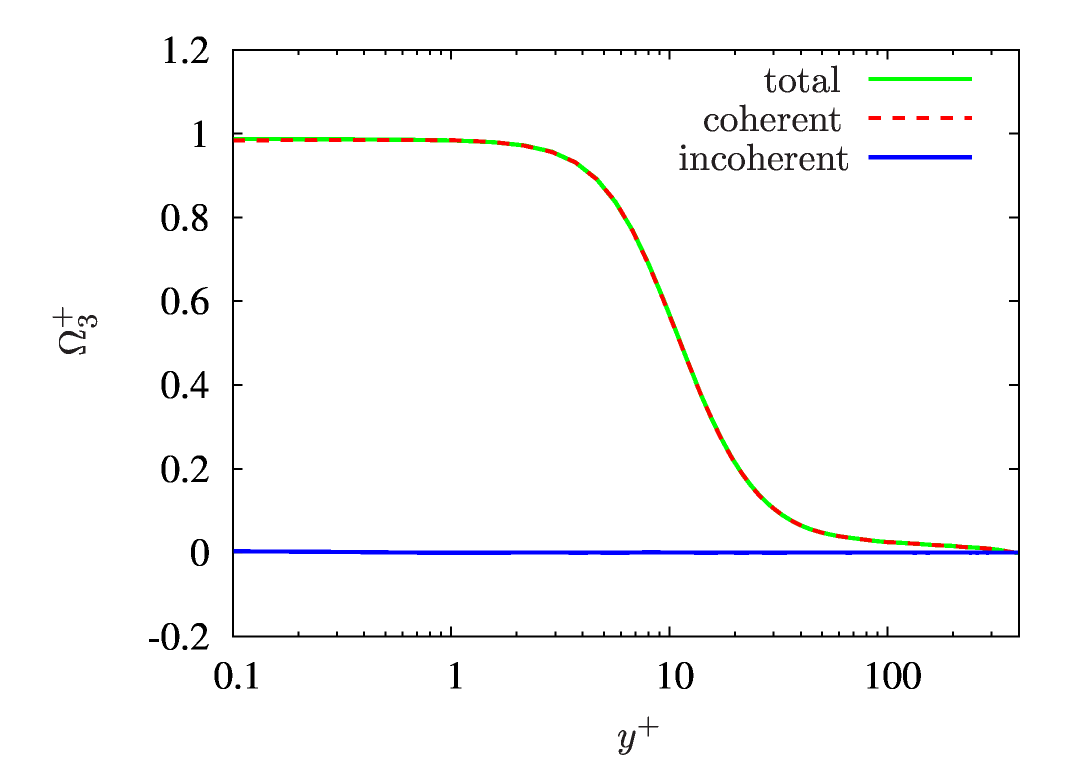}
\caption{Streamwise mean velocity $U_1^+(y^+)$ (left) and spanwise mean vorticity $\Omega_3^+(y^+)$ for total, 
coherent and incoherent flows, together with the DNS results of Ref. ~\cite{MKM} in the lin-log representation.}
\label{meanvor}
\end{figure*}

We analyze the statistics of the mean velocity and vorticity profiles of the coherent and incoherent flows, 
and compare them with the total flow.
The results are averaged over 40 snapshots.
Figure \ref{meanvor} shows the $y^+$-dependence of the streamwise mean velocity $U_1^+(y^+)$ and of the spanwise mean vorticity 
$\Omega_3^+(y^+)$, averaged over $x_1$-$x_3$ planes, for the total, coherent and incoherent flows. 
It is observed that the coherent flow perfectly preserves $U_1^+(y^+)$ and $\Omega_3^+(y^+)$,  
while both incoherent contributions are very weak.
It can be noted that $U_2^+(y^+)$ vanishes identically and that $U_3^+(y^+)$ almost vanishes for the total, coherent and incoherent flows.
This implies that $\Omega_1^+(y^+)$ is almost zero, and $\Omega_2^+(y^+)$ is identically zero. 
The comparison of $U_1^+$ with the DNS data at $Re_\tau=395$ in Moser {\it et al}.~\cite{MKM} confirms the validity of the present DNS.

\subsection{Statistics of velocity and vorticity fluctuations}

\begin{figure*}[tb]
\includegraphics[width=80mm]{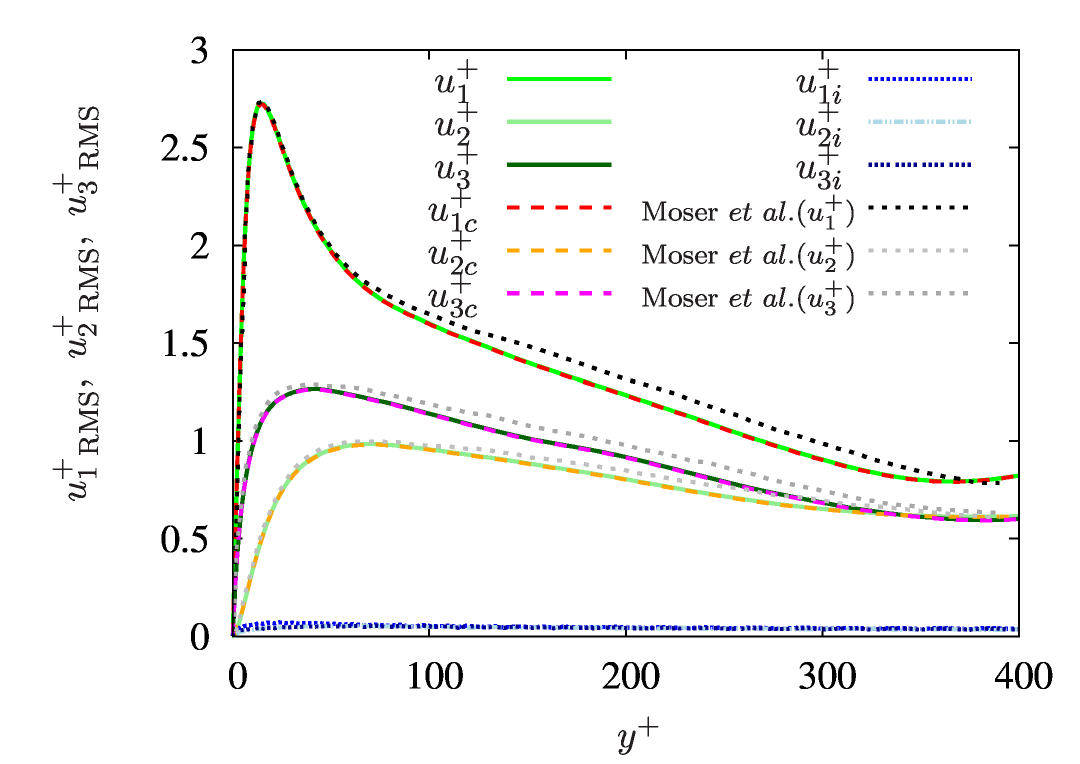}
\includegraphics[width=80mm]{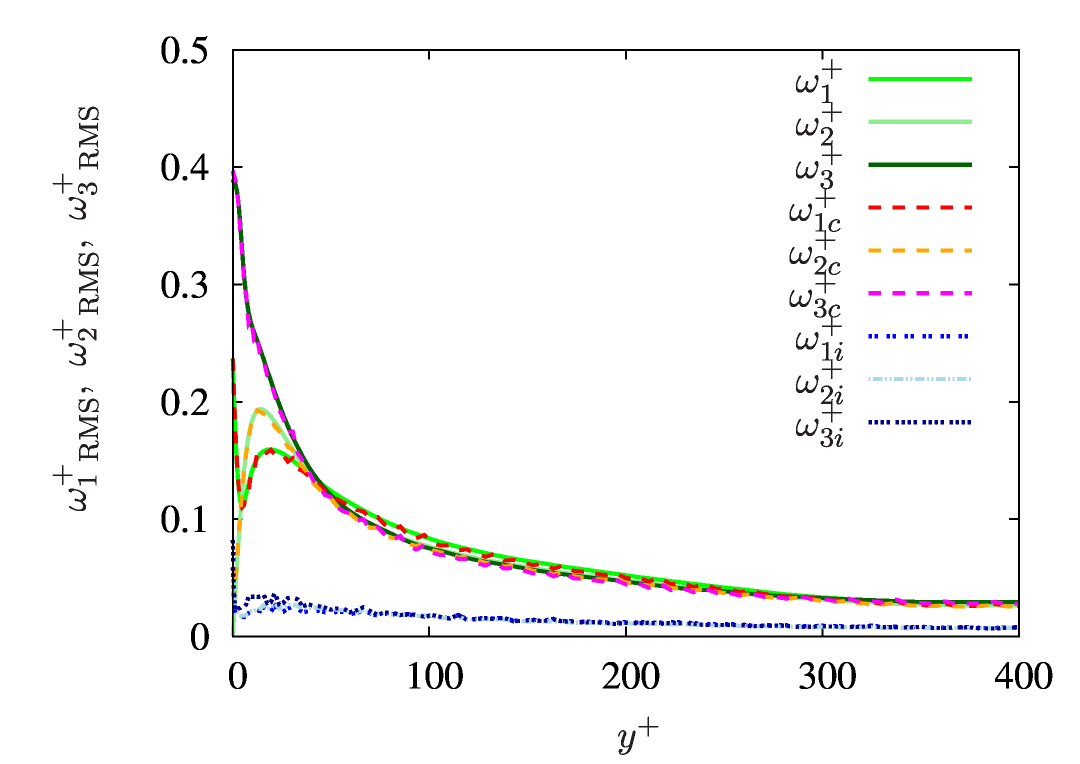}
\caption{RMS of $u_j^+$ (left) and RMS of $\omega_j^+$ (right) for total, coherent and incoherent flows. }
\label{RMS_middle}
\end{figure*}

The root-mean-square (RMS) of the velocity fluctuations $u_j^+$ $(j=1,2,3)$ as a function of $y^+$ are shown in Fig. \ref{RMS_middle} (left).
Again, we find an excellent agreement between the total and the coherent flow for all values of $y^+$, 
while the incoherent contribution is negligibly small.
For RMS of the vorticity fluctuations $\omega_j^+$ in Fig. \ref{RMS_middle} (right), 
the coherent contributions well preserves the total RMS of $\omega_j^+$.
The vorticity RMS of the incoherent flow is much smaller than that of the total flow.

\subsection{Probability density functions of velocity and vorticity}

\begin{figure*}[tb]
\begin{tabular}{c|c}
\begin{minipage}{0.5\textwidth}
\centering
\includegraphics[width=80mm]{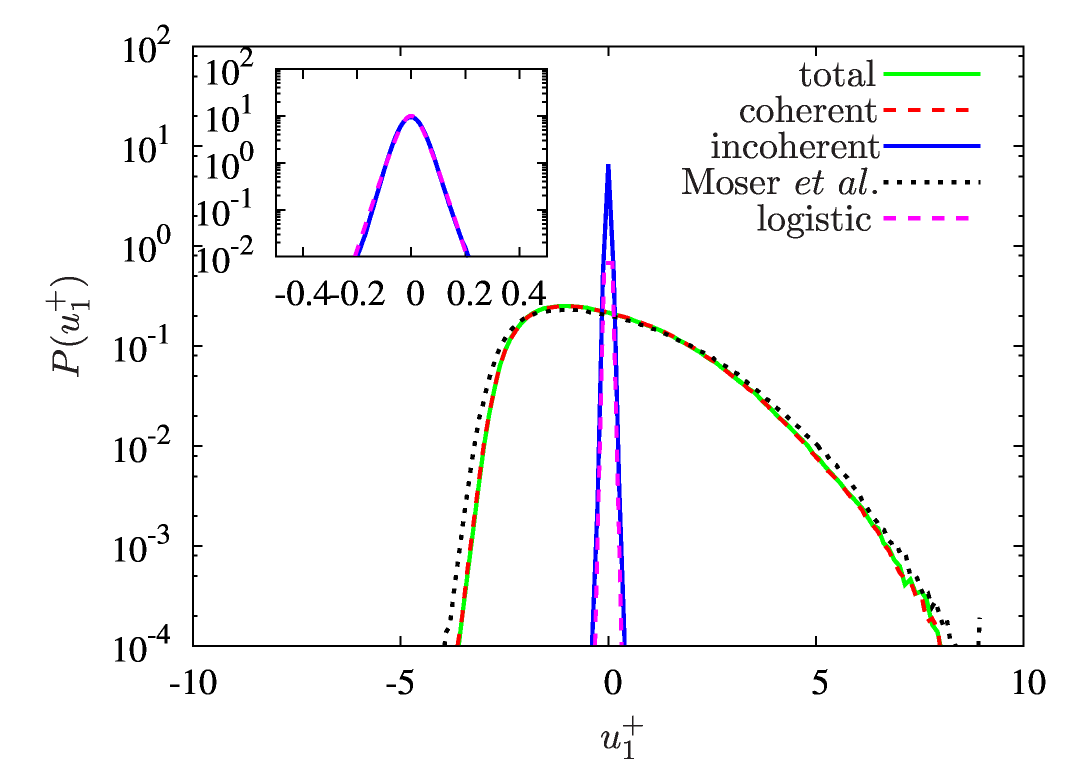}\\
\includegraphics[width=80mm]{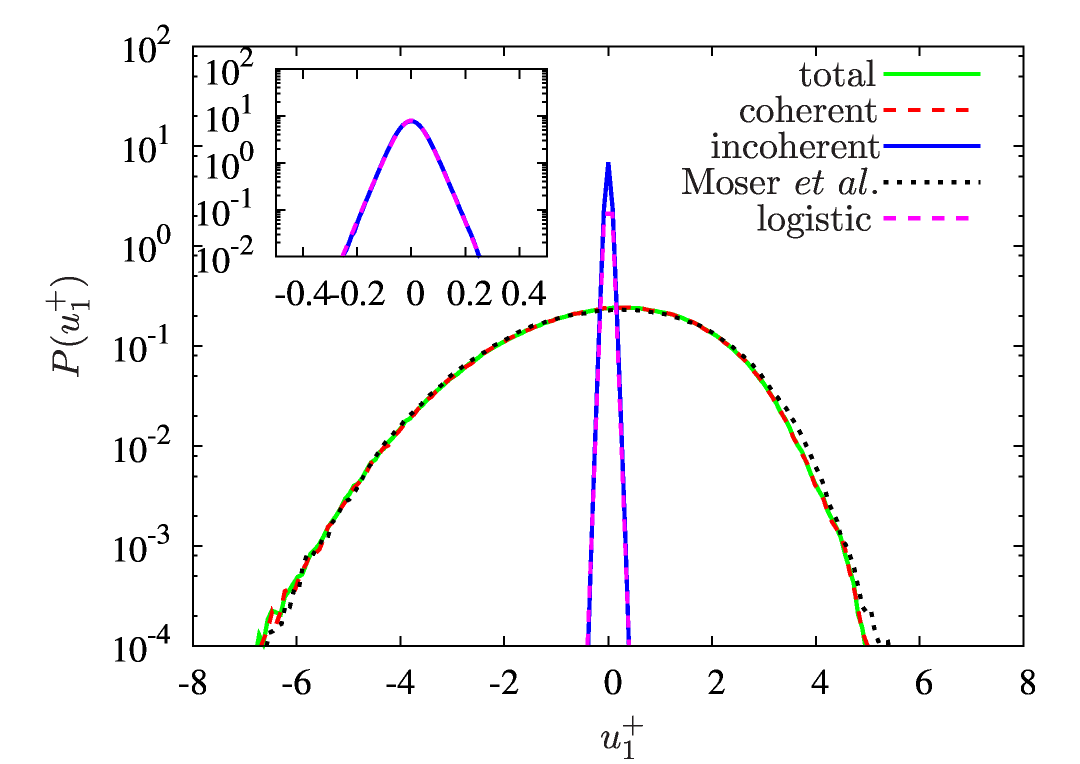}\\
\includegraphics[width=80mm]{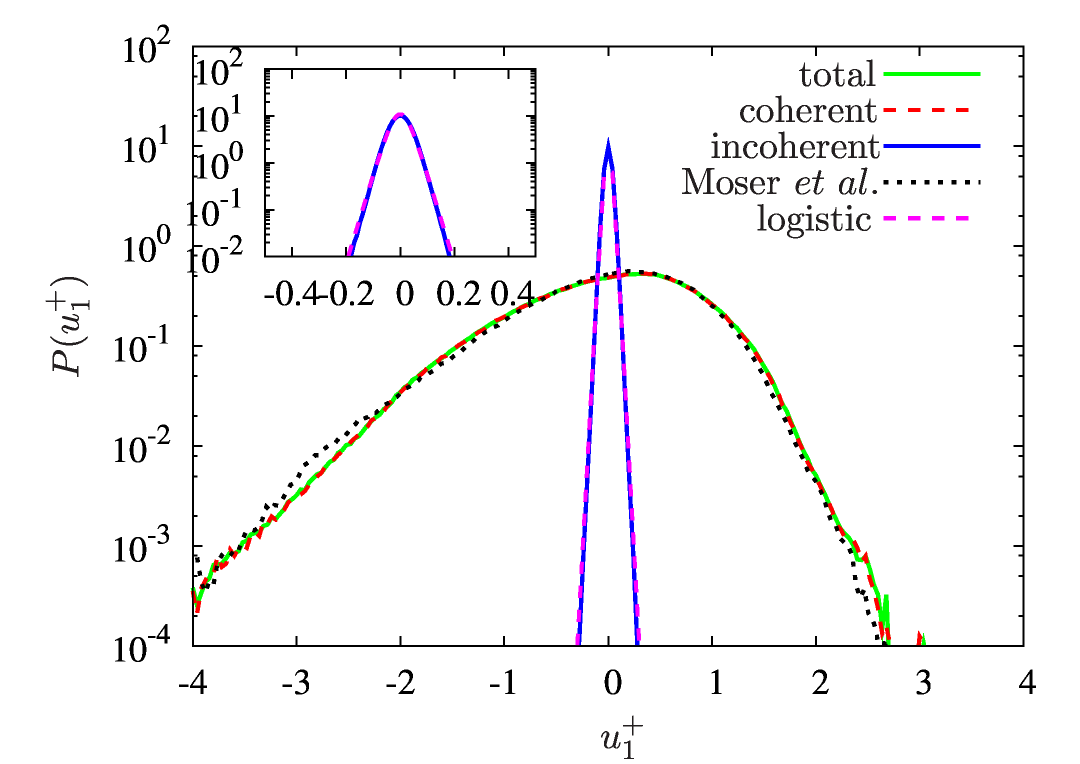}
\end{minipage}
&
\begin{minipage}{0.5\textwidth}
\centering
\includegraphics[width=80mm]{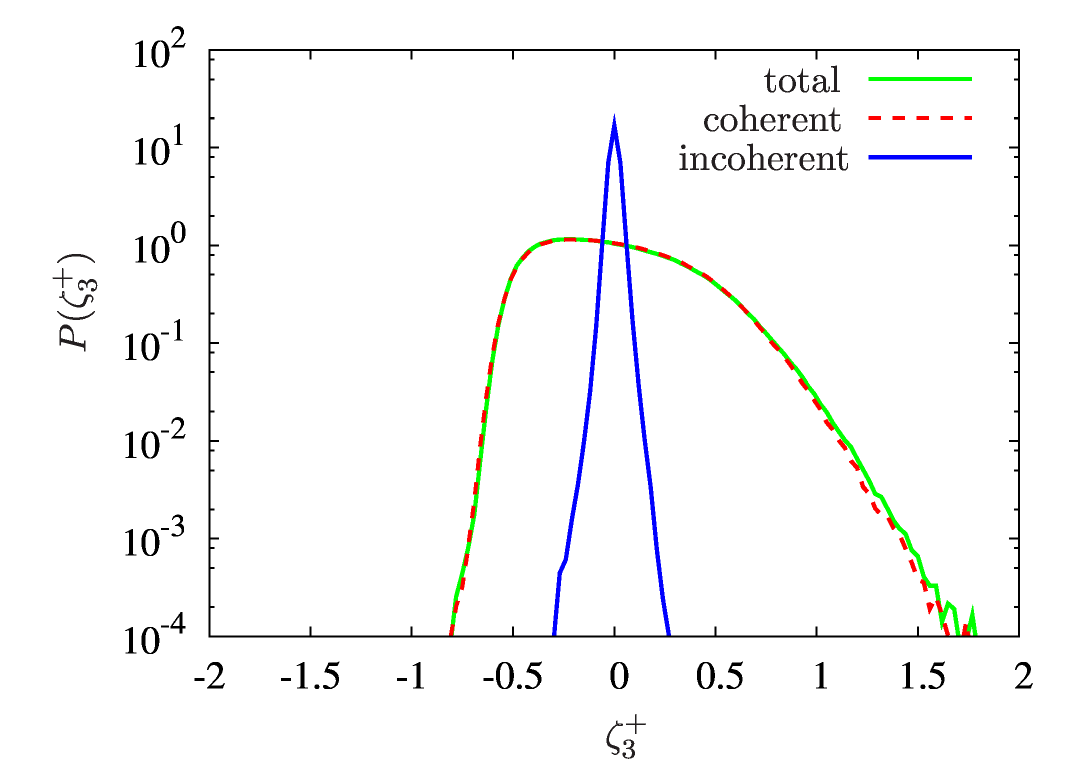}\\
\includegraphics[width=80mm]{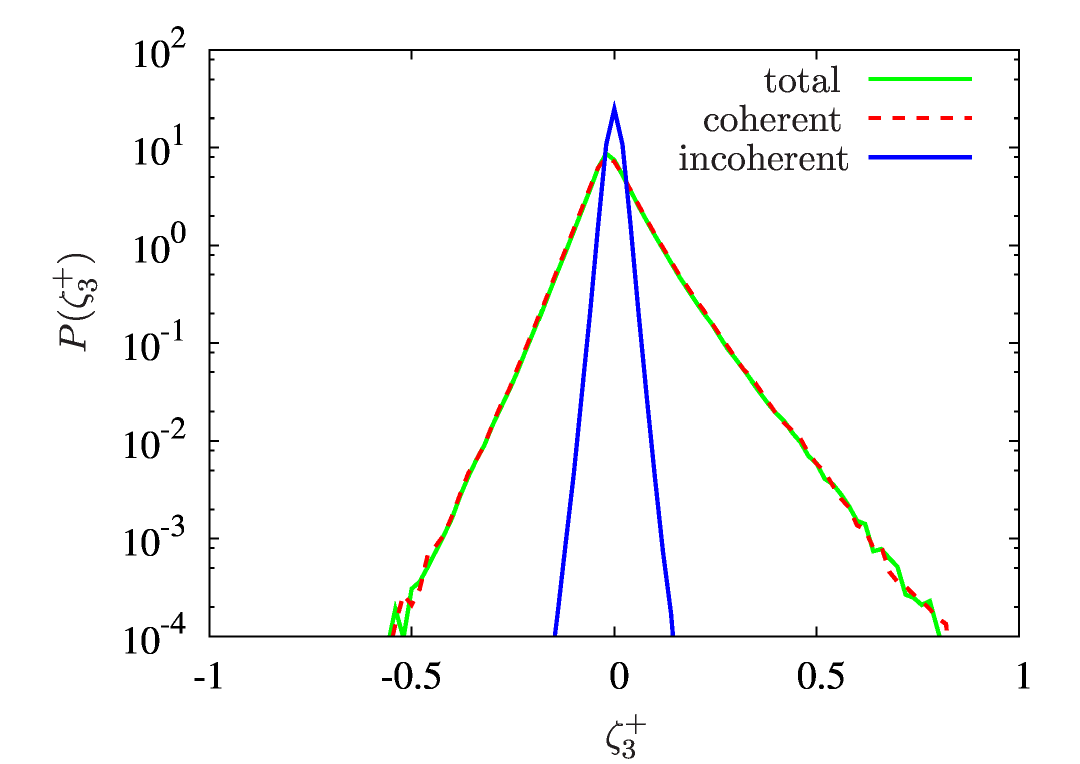}\\
\includegraphics[width=80mm]{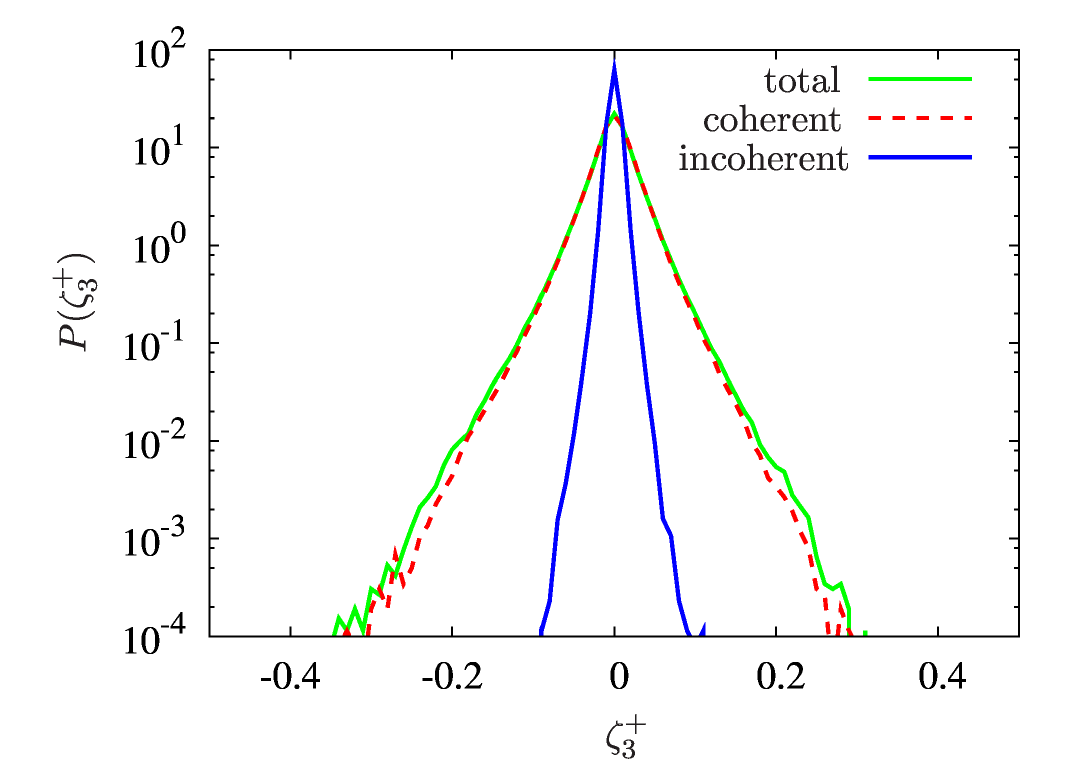}
\end{minipage}
\end{tabular}
\caption{PDFs of $u_1^+$; (top left) $y^{+}=4.6$ viscous sublayer, (top right) $y^{+}=96.8$ around the log region, 
and (bottom) $y^{+}=378.8$ around the center of the channel. }
\label{PDF}
\end{figure*}

Figure \ref{PDF} (left) shows the probability density functions (PDFs), estimated using histograms with $200$ bins, 
of the streamwise velocity fluctuations $u_1^+$ for the total, coherent and incoherent flows  at three representative positions $y^+$: 
in the viscous sublayer, the log region, and near the center of the channel.
In all cases, we observe that the PDFs for the total and coherent velocity fluctuations perfectly superimpose, 
which indicates that high order statistics are well preserved by the coherent flow.
We also find that the velocity PDFs remain skewed in the different regions and agree well with the data of Ref.~\cite{MKM}, using appropriate renormalization.
In contrast, the PDFs of the incoherent velocity fluctuations are symmetric, and have strongly reduced variances.
For the incoherent velocity, we analyzed $y^+$-dependent flatness, 
and found values around 4 in the viscous sublayer and in the log region, 
which decrease to 3.6 near the center of the channel.
For the $y^+$-dependent skewness, fluctuations around zero are observed with an amplitude below 0.05.
The PDFs of the incoherent velocity well superimpose the logistic distribution with zero mean and the variances $\sigma^2(y^+)$ of the incoherent velocity, 
though their flatness is 1.2, which is much smaller than the PDFs of the incoherent velocity.
The distributions $P(u_1^+)$ are given by $\exp (-\pi u_1^+ /s)/\{ s(1+\exp (-u_1^+ /s) \}$, where $s=3^{1/2} \sigma(y^+)/\pi$.

In Fig. \ref{PDF} (right), we illustrate the PDFs of the vorticity fluctuations $\omega_3^{+}$ at three representative positions $y^+$: 
in the viscous sublayer, the log region, and near the center of the channel. 
The coherent vorticity fluctuations well represent the total vorticity PDFs which are skewed in all cases, 
while the corresponding incoherent PDFs are symmetric.
The variances of these incoherent PDFs are respectively significantly weaker than the variances of the total and coherent PDFs. 

\subsection{Energy spectra}

\begin{figure*}[tb]
\includegraphics[width=80mm]{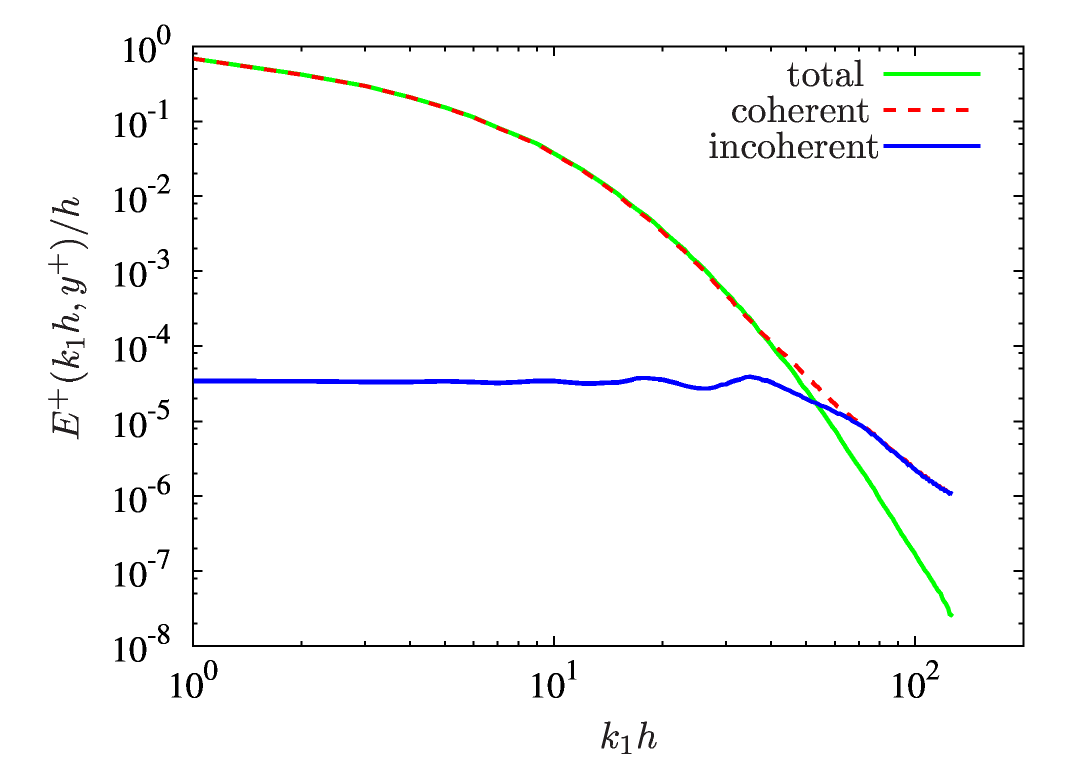}
\includegraphics[width=80mm]{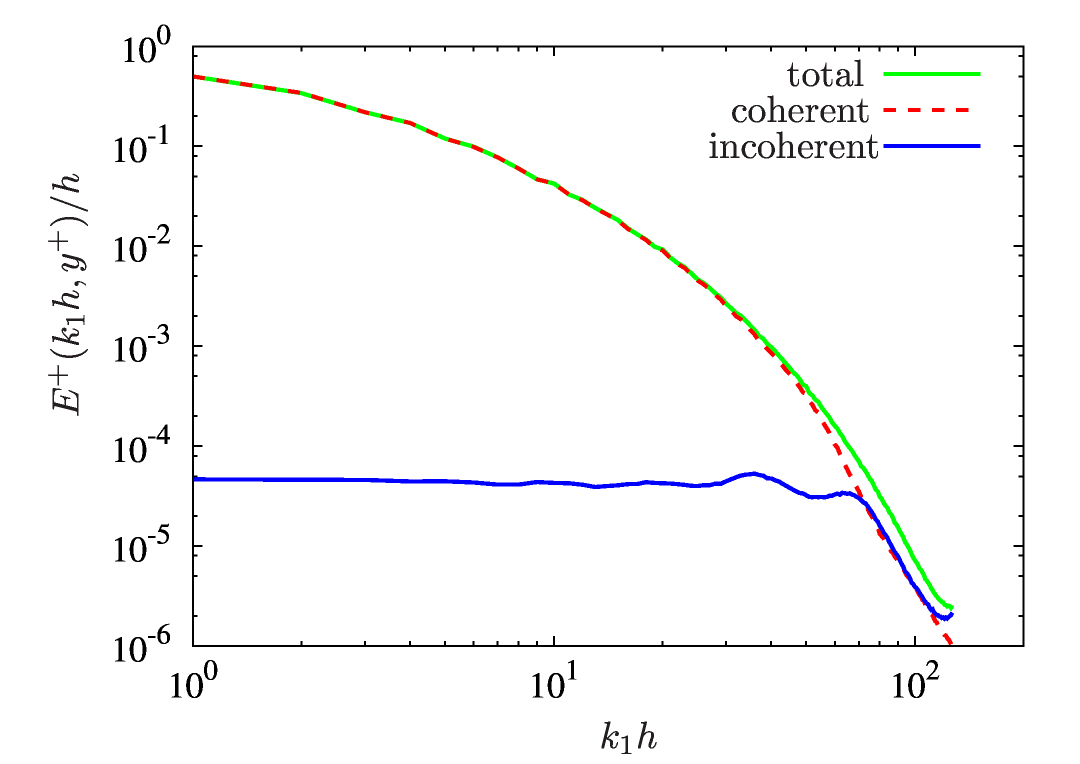}
\includegraphics[width=80mm]{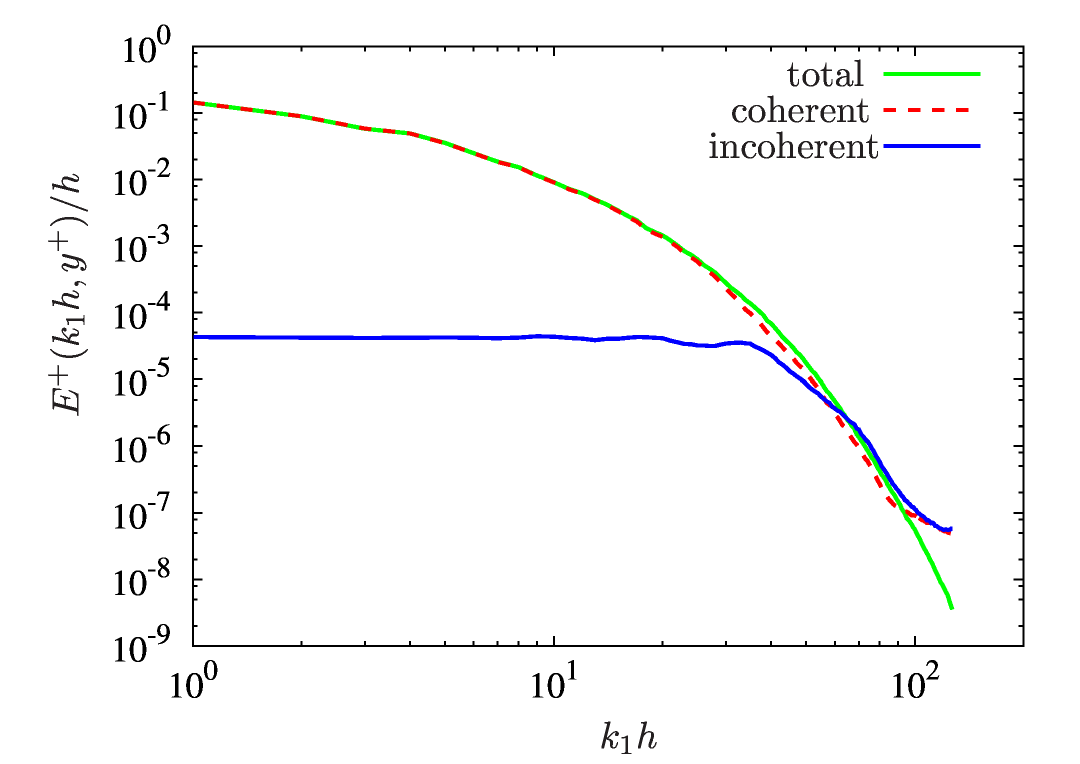}
\caption{Dimensionless energy spectra of $u_1^+$ in the $x_1$-direction at three representative $y^+$:
(top left) $y^+=4.6$ viscous sublayer, (top right) $y^+=96.8$ around the log region, and (bottom) $y^+=378.8$ around the center of the channel. }
\label{spectra}
\end{figure*}

To get insight into the scale distribution of turbulent kinetic energy we analyze the
1D energy spectra of the streamwise velocity $u_1^+$ in the streamwise direction $E^+(k_1 h,y^+)$, 
which is defined as
$E^+(k_1 h,y^+) = \sum' |{\hat u}_i(k_1 h,y^{+},k_3 h)|^2 /2$, 
where ${\hat u}_i(k_1 h,y,k_3 h)$ is the Fourier transform of ${\bm u}^+({\bm x})$ in the $x_1$-$x_3$ planes,
$\sum'$ denotes the summation in terms of all $k_3$.
The results are shown in Fig.~\ref{spectra}, again for the total, coherent an incoherent flows at three representative positions; 
in the viscous sublayer, the log layer and near the center of the channel.
The dimensionless wavenumber in the $x_1$-direction is denoted by $k_1 h$. 
Figure \ref{spectra} shows that the spectral distribution of turbulent kinetic energy is well preserved by the coherent flow, 
from the viscous sublayer to the center of the channel.
In contrast, the incoherent energy exhibits an almost flat spectrum,
which corresponds to equipartition of incoherent energy, 
i.e., decorrelation of the incoherent flow in physical space.

At large wavenumbers close to the cut-off scale, imposed by the resolution of the DNS,
we find that the incoherent energy dominates the total energy in the viscous sublayer and the log layer, 
while it dominates the coherent energy around the center of the channel.
However, this is not surprising since the wavelet decomposition is orthogonal for vorticity but not for velocity, 
due to the fact that the Biot-Savart operator is not diagonal in wavelet space.
Note that ${\hat {\bm \omega}}_c^+ (k_1,y^+,k_3)$
and ${\hat {\bm \omega}}_i^+ (k_1,y^+,k_3)$ are not orthogonal for any fixed $(k_1,k_3)$ at every $y^+$.
But even though the cross-term $\langle {\bm \omega}_c^{+} {\bm \cdot} {\bm \omega}_i^{+} \rangle(x_2) \neq 0$, 
its averaged contribution vanishes, $\int_{-h}^h d x_2 \langle {\bm \omega}_c^+  {\bm \cdot} {\bm \omega}_i^+ \rangle(x_2) =0 $.

\begin{figure}[tb]
\begin{center}
\includegraphics[scale=0.8]{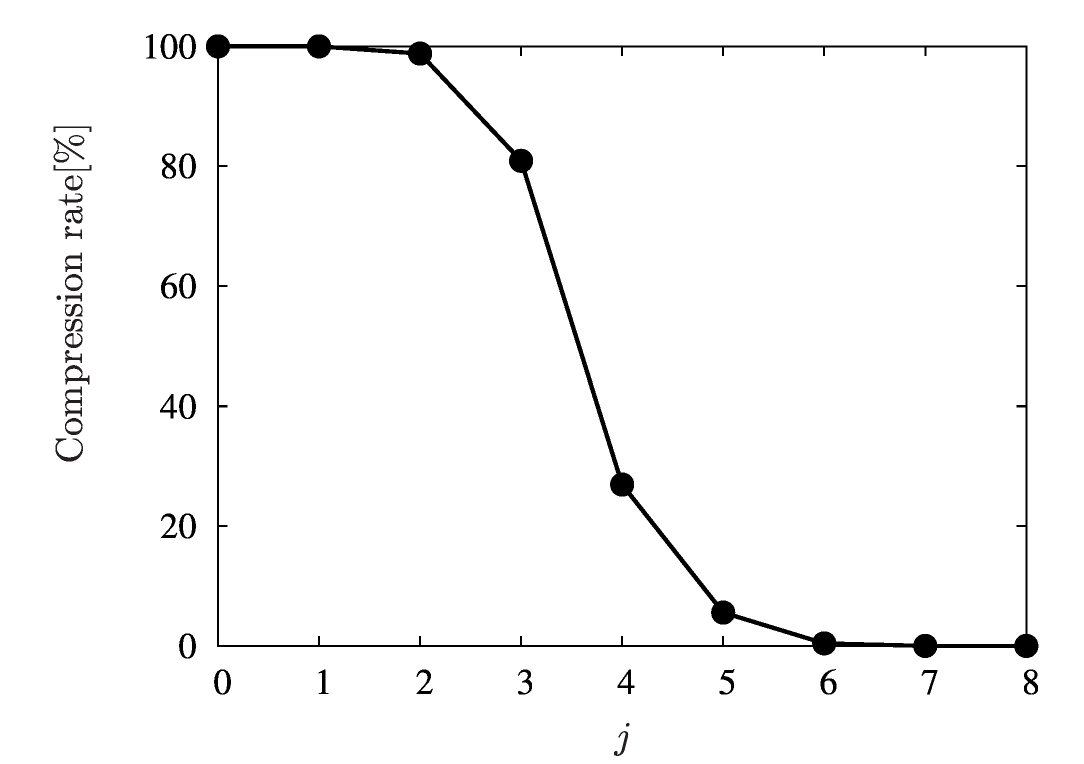}
\caption{Scale-dependent compression rate.}
\label{cr_scale}
\end{center}
\end{figure}

The compression is most efficient for small scales, i.e., small $j$ and large wavenumbers (Fig. \ref{cr_scale}).
This implies that the scale-by-scale incoherent enstrophy is comparable or larger than the scale-by-scale coherent enstrophy.

\subsection{Nonlinear dynamics}

\begin{figure*}[tb]
\includegraphics[width=80mm]{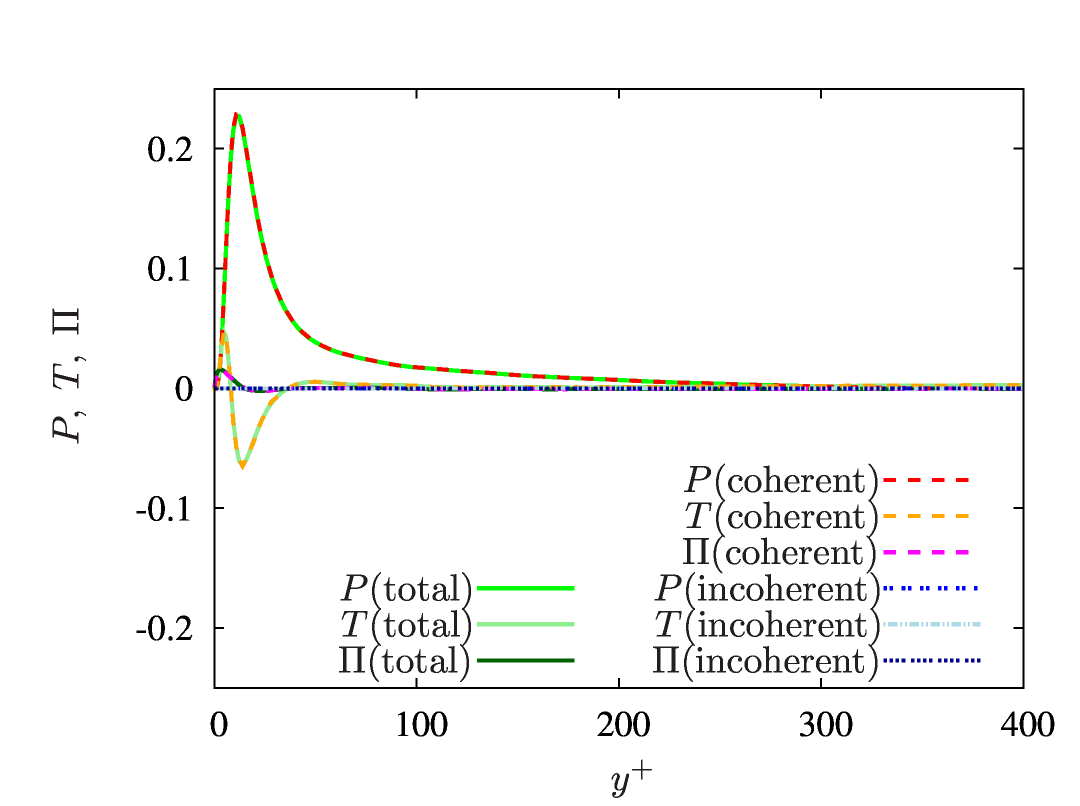}
\includegraphics[width=80mm]{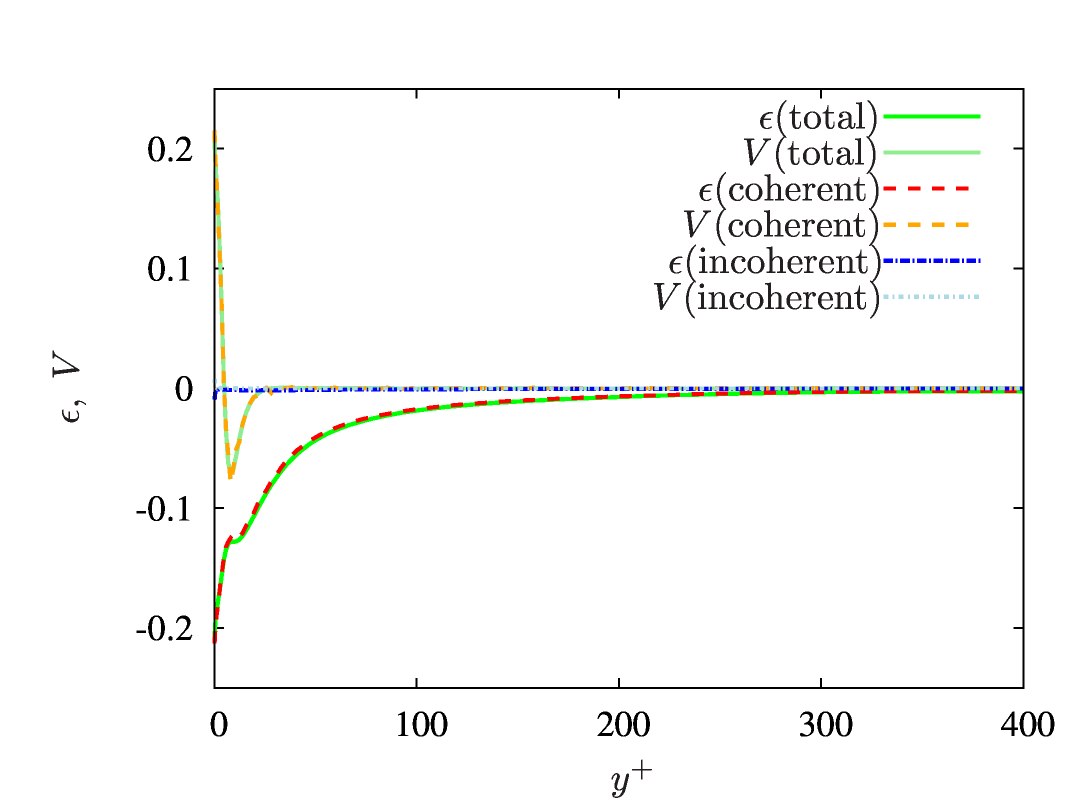}
\caption{Production term $P$, turbulent diffusion term $T$ and pressure diffusion term $\Pi$ vs. $y^+$ (left).
Energy dissipation $\epsilon$ and viscous diffusion $V$ vs. $y^+$ (right). Coherent and incoherent contributions are presented together with the total one.}
\label{Energy_budget_m}
\end{figure*}

\begin{figure*}[tb]
\includegraphics[width=80mm]{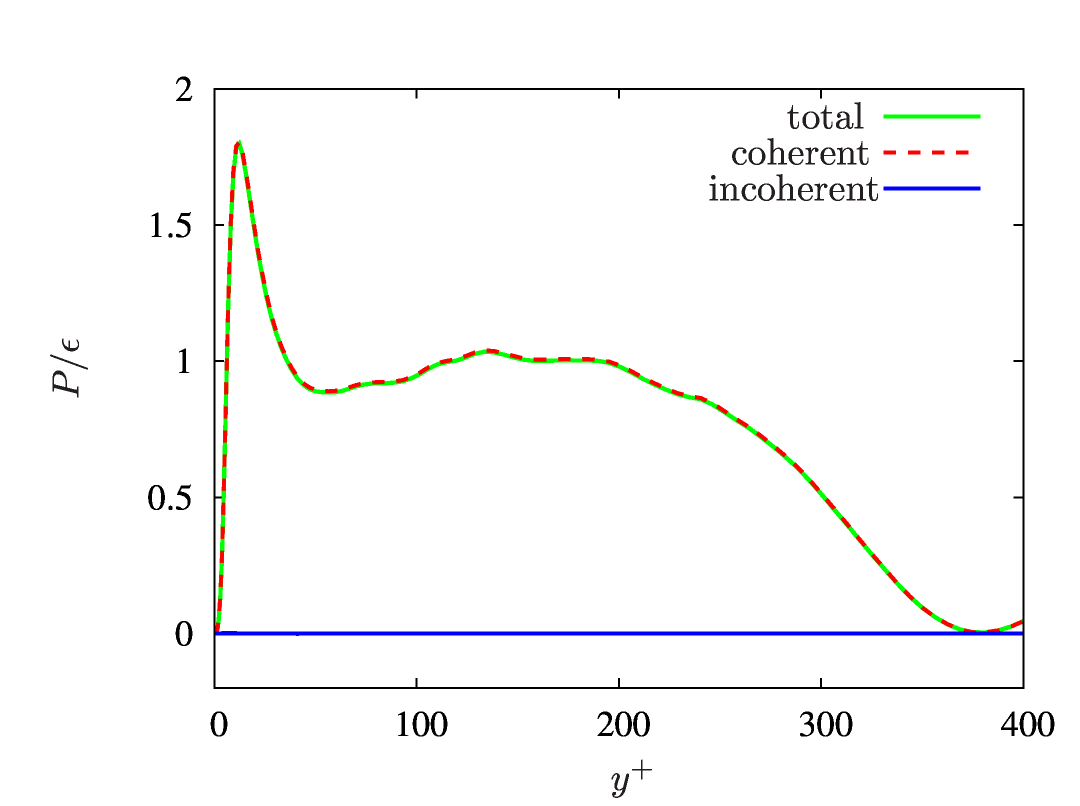}
\includegraphics[width=80mm]{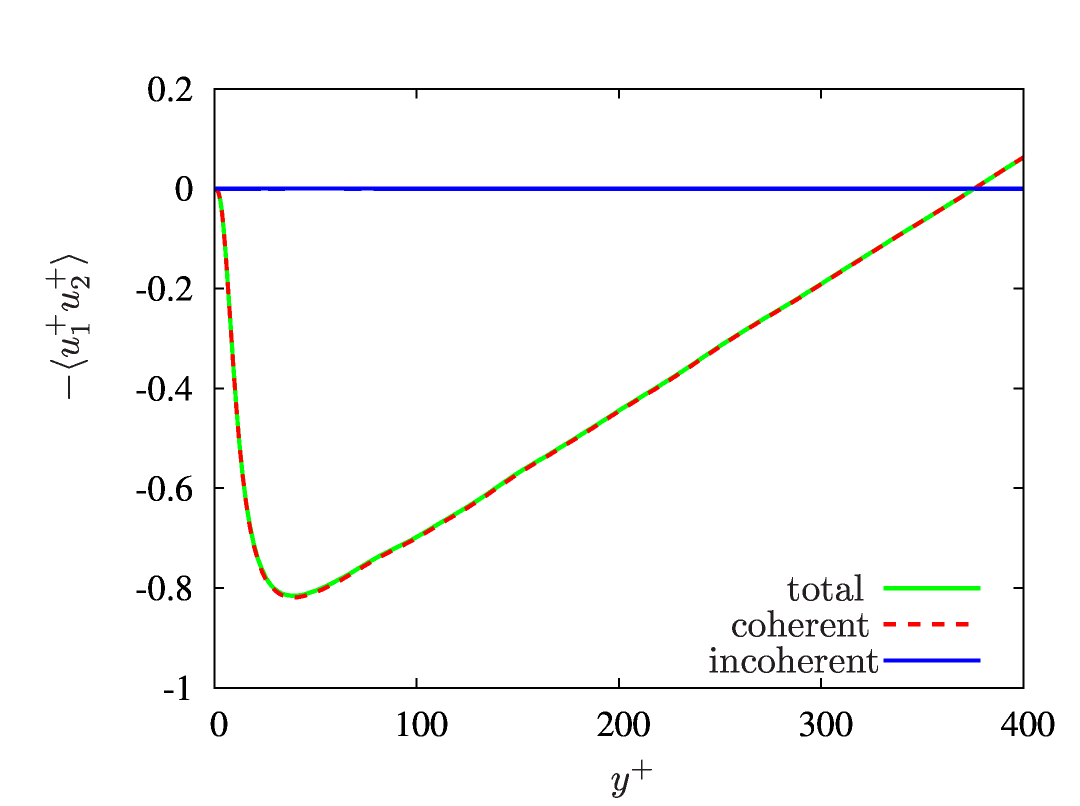}
\caption{The ratios of total, coherent and incoherent productions $P$ to the dissipation of turbulent kinetic energy for total flow, 
$\epsilon$, vs. $y^+$(left). The Reynolds stress $-\langle u_1^+ u_2^+ \rangle$ vs. $y^{+}$(right).}
\label{Povere}
\end{figure*}

To get further insight into the nonlinear dynamics,
we consider the energy budget given in the equation for $\langle u_j^+ u_j^+ \rangle/2$ per unit mass~\cite{MasourKimMoin}:
\begin{equation}
\begin{split}
& \frac{1}{2} \left( \partial_t + U_j^+ \partial_j \right) \left< u_j^{+} u_j^{+} \right> = \\
& P({\bm v}^+) + T({\bm u}^+) + \Pi({\bm u}^+,p^+) - \epsilon({\bm u}^+) + V({\bm u}^+), \label{En_budget}
\end{split}
\end{equation}
where
$P({\bm v}^+) = - \left< u_j^{+} u_l^{+} \right> \partial_l U_j^{+}$, 
$T({\bm u}^+) = -  \partial_l \left< u_j^{+} u_j^{+} u_l^{+} \right>/2$, 
$\Pi({\bm u}^+,p^+) = - \partial_l \left< p^{+} u_l^{+} \right>$, 
$\epsilon({\bm u}^+) = \nu \left< \partial_l u_j^{+} \partial_l u_j^{+} \right>$, 
$V({\bm u}^+) = \nu \partial_l \partial_l \left< u_j^{+} u_j^{+} \right>$.
In Fig. \ref{Energy_budget_m} (left), 
we see that three nonlinear coherent contributions, 
corresponding to production $P({\bm v}_c^+)$, turbulent diffusion $T({\bm u}_c^+)$ and pressure diffusion $\Pi({\bm u}_c^+,p^+_c)$, 
are in good agreement with the corresponding total ones.
Hence, the coherent flow almost perfectly preserves the nonlinear dynamics.
Thus, we anticipate that the departure of the coherent flow from the total flow is negligibly small.
Indeed, the incoherent contribution to the different terms, defined by $P({\bm v}^+)-P({\bm v}_c^+)$, $T({\bm u}^+)-T({\bm u}_c^+)$ and 
$\Pi({\bm u}^+,p^+)- \Pi({\bm u}_c^+,p^+_c)$, is almost zero.
The two viscous contributions, $\epsilon({\bm u}^+)$ and $V({\bm u}^+)$, 
are also well retained by the coherent flow, $\epsilon({\bm u}_c^+)$ and $V({\bm u}_c^+)$, 
as confirmed in Fig. \ref{Energy_budget_m} (right).
In the viscous sublayer, the incoherent flow has small contribution on $\epsilon({\bm u}^+)$ and $V({\bm u}^+)$.
The incoherent contribution to the viscous terms, respectively measured by $\epsilon({\bm u}^+)-\epsilon({\bm u}_c^+)$ and $V({\bm u}^+)-V({\bm u}_c^+)$, 
becomes even smaller and more negligible as $y^+$ increases, a behavior which is expected.

The ratio between the production and the dissipation yields insight into the equilibrium of the turbulent flow in the log region,
as discussed in Ref.~\cite{MKM}.
Figure \ref{Povere} (left) shows this balance.
Considering $P({\bm v}_c^+) / \epsilon({\bm u}^+)$, the coherent contribution perfectly superimposes with the ratio of the total flow, 
$P({\bm v}^+) / \epsilon({\bm u}^+)$.
The corresponding quantity for the incoherent flow, $\{P({\bm v}^+)-P({\bm v}_c^+) \}/ \epsilon({\bm u}^+)$,
is negligible,
as expected from Fig. \ref{Energy_budget_m} (left).

The Reynolds stress defined by $-\langle u_1^+ u_2^+ \rangle$ measures the fluctuation of turbulent momentum.
The analysis of the Reynolds stress provides detailed information on the contribution to the turbulence production 
from various events occurring in the flows.
Figure \ref{Povere} (right) shows that 
the coherent Reynolds stress well represents the Reynolds stress for the total flow, 
while its incoherent contribution is negligibly small.
The interaction between the coherent flows is predominant over the stress.
In contrast, the remaining interactions play a non-significant role in the stress, 
not only between the incoherent flows but also between the coherent and the incoherent flows. 

\section{\label{sec4}Conclusions and perspectives}

DNS data of turbulent channel flow at moderate Reynolds number have been analyzed using the coherent vorticity extraction method.
Boundary adapted isotropic wavelets have been developed and implemented into a fast wavelet transform.
By thresholding the wavelet coefficients, the flow has been decomposed into coherent and incoherent contributions.
We found that few percent of wavelet coefficients, i.e., 6 \%, are sufficient to represent the coherent structures of the flow.
The low order statistics, mean velocity, mean vorticity, RMS velocity and RMS vorticity of the coherent part agree very well with those of the total flow.
A spectral decomposition of turbulent kinetic energy confirms that the coherent flow matches the spectrum all along the inertial range.
In contrast, the incoherent flow exhibits energy equipartition which suggests that filtering it out corresponds to modeling turbulent dissipation.
In order to obtain reliable statistical results, averaging over 40 flow snapshots has been performed.
To get insight into the flow dynamics we analyzed the energy budget and we found that the coherent flow almost perfectly retains the nonlinear dynamics.
The production/dissipation ratio of the coherent flow superimposes well the one of the total flow in the log layer, 
while the interactions between incoherent-incoherent and coherent-incoherent contributions are negligibly small.
Although the coherent and incoherent vorticity fields are not perfectly divergence free. 
The divergence issue is not crucial as discussed in appendix \ref{app_div}.

The present construction requires that the DNS data be interpolated onto an equidistant grid.
This limits the applicability of the current CVE algorithm as higher resolution DNS data
may not be handled due to the implied memory requirements.
One way to overcome this is the use of Chebyshev wavelets, see e.g.~\cite{FU,Plonka}.
In the appendix \ref{Chewave}, we tested this approach and we have shown that similar results in terms of statistics and compression rate are indeed obtained.

The CVE results are encouraging for developing coherent vorticity simulation (CVS) of wall bounded turbulent flows.
We anticipate that for higher Reynolds number the compression rate will further improve, similar to what was found for isotropic turbulence~\cite{OYSFK07}.
CVS is based on a deterministic computation of the coherent flow evolution using an adaptive orthogonal wavelet basis ~\cite{FS01}.
The influence of the incoherent background flow is neglected to model turbulent dissipation.
Applications of CVS to turbulent mixing layers and isotropic turbulence can be found in Refs.~\cite{SFPR05jfm} and ~\cite{SIAMMMS}, respectively.

Some challenges for future work are that the wavelet bases are not orthogonal in 2D planes for fixed $ y^+$.
This implies that 2D statistics cannot be done, especially at small scales.
In this case it would be better to apply 2D wavelets in each plane, as done in previous work \cite{DM1}.

\begin{acknowledgments}
We thank Professor Javier Jim\'enez for fruitful discussions about computing the mean pressure and the divergence issue.
The computations were carried out on the FX100 systems at the Information Technology Center of Nagoya University.
This work was partly supported by JSPS KAKENHI Grant Numbers (S) 24224003 and (C) 25390149.
K.S. thankfully acknowledges financial support from ANR, contract SiCoMHD (ANR-Blanc 2011-045).
M.F. and K.S. acknowledge support by the French Research Federation for Fusion Studies within the framework of
the European Fusion Development Agreement (EFDA).
\end{acknowledgments}

\appendix
\section{\label{app_div}Divergence issues}

\begin{figure*}[tb]
\includegraphics[width=80mm]{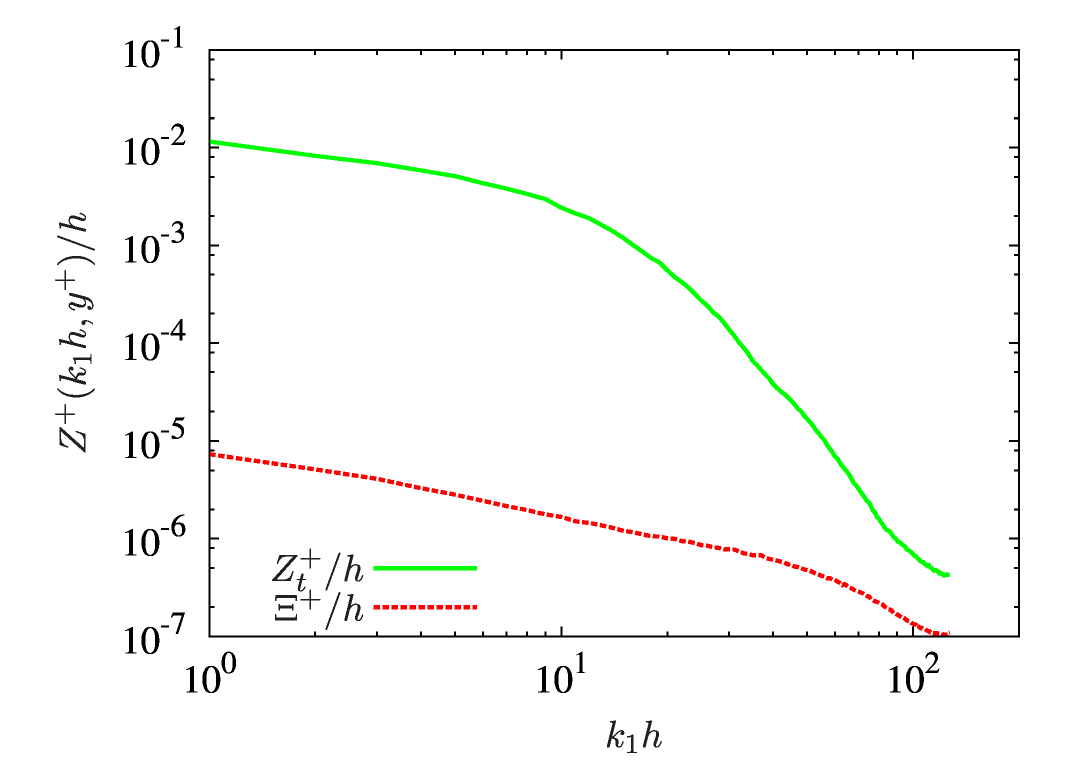}
\includegraphics[width=80mm]{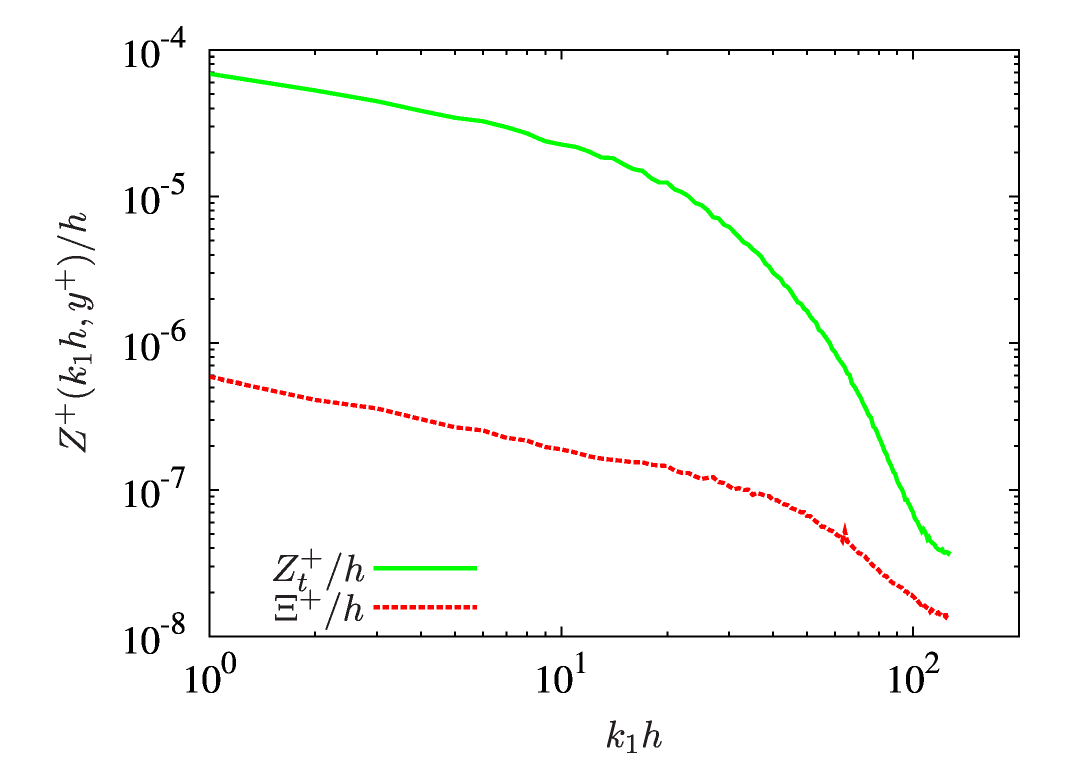}
\caption{Enstrophy spectra, $Z^+(k_1h,y^+)$, and spectra of the divergence component $\Xi^+(k_1h,y^+)$ in the $x_1^+$-direction, 
at (left) $y^+=4.6$ in the viscous sublayer, and (right) at $y^+=381.9$ around the center of the channel. }
\label{Div}
\end{figure*}

The vector-valued wavelet basis used here is not divergence-free,
since the orthogonal wavelet transform does not commute with the differential operator. 
Thus, the coherent vorticity, ${\bm {\omega}}_c$, and also the incoherent one, ${\bm {\omega}}_i$, are not divergence-free. 
In the following, we quantify the $y^+$-dependent contribution of the divergent component $\nabla \xi$ of ${\bm {\omega}}_c$ 
on the streamwise vorticity spectra in the $x_1$-direction.
Figure \ref{Div} shows dimensionless spectra of total streamwise vorticity ${\bm \omega}^+$ and those of $\nabla \xi^+$ 
at two representative values of $y^+$, which are respectively located near the wall and around the center of the channel. 
The contribution of $\nabla \xi^+$ appears mostly in the dissipative range, not only in the viscous sublayer but also around the center of the channel.
It can be seen that the contributions of $\xi^+$ are weak in the lower wavenumber region. 
The intensity of $\nabla \xi^+$, denoted by $\langle |\nabla \xi|^2 \rangle (y^+)$, is about $2.8 \times 10^{-2}\%$ of the total enstrophy in the viscous sublayer, 
and about $1.89\%$ around the center of the channel.
Therefore, this divergence issue in ${\bm \omega}_c$ is negligible for the statistics, but also for simulations, since ${\bm \omega}_c$ is almost divergence-free.

\section{\label{Chewave}CVE using Chebyshev wavelets}

In the following, we briefly summarize Chebyshev wavelets 
which yield an alternative construction of wavelets on the interval~\cite{Maday}.
The idea is to perform a change of variables, similar to what is done for the trigonometric definition of Chebyshev polynomials. 
The efficient numerical implementation of Chebyshev wavelets is based on the periodic wavelet transform, in analogy 
with the fast Chebyshev transform which uses the cosine transform. 
The CVE results presented here use Chebyshev wavelets in the $x_2$-direction instead of the CDJV wavelets, 
while in the $x_1$ and $x_3$-directions periodic Coiflet 30 wavelets are used. 

\subsection{On Chebyshev wavelets}

Using the coordinate transform  $x=\cos(\theta)$ we map the interval $x \in [-1,1] $ onto  $\theta \in [0,\pi]$.
Then $\pi$-periodic orthogonal wavelets $\psi^P (\theta)$ are used to construct wavelets $\psi^B(\theta)$, \cite{FU}, which are even functions:
\begin{equation}
\psi^B (\theta) = \psi^P (\theta) + \psi^P (\pi - \theta).
\end{equation}
The corresponding dilated and translated wavelets are obtained by $\psi^B_{j,i} (\theta) = 2^{j/2} \psi^B (2^j \theta - i)$.
Setting $\theta = \arccos x$ we obtain the boundary wavelets $ \psi^B (x)$ on the interval $[-1, 1]$ which 
yield an orthogonal basis with respect to the weighted scalar product, i.e., \\
$$ \int_{-1}^{1} \, \psi_{j,i} ^B (x)\psi^B_{j',i'} (x) /(1-x^2)^{1/2} \, dx =   \delta_{jj'} \, \delta_{i i'} \, .$$

To compute the Chebyshev wavelet transform efficiently we use periodic orthonormal Coiflet 30 wavelets with period $2 \pi$ 
and extend the vorticity ${\bm \omega} (x_1, \theta, x_3)$ as an even function ${\bm g}(x_1,\theta, x_3)$ for each $(x_1,x_3)$,
\begin{equation}
{\bm g}(x_1,\theta, x_3) = \left\{
\begin{array}{ll}
{\bm \omega} (x_1,\theta,x_3) &{\mathrm{for}} \,\,\,  0 \le \theta \le \pi,\\
{\bm \omega} (x_1,-\theta,x_3) & {\mathrm{for}} \,\,\, -\pi \le \theta <0.
\end{array}
\right.
\end{equation}
Before applying the extension of ${\bm \omega}$, we interpolate the vorticity given on $192$ Chebyshev grid points 
onto $256$ equidistant grid points in the $\theta$-coordinate.
Then we can proceed with the CVE method and apply the fast wavelet transform to ${\bm g}$ 
using 3D orthogonal wavelets  constructed by a tensor product from $\psi^P(x_1)$, $\psi^P (\theta)$ and $\psi^P(x_3)$.

\subsection{Numerical results}

\begin{figure}[tb]
\begin{center}
\includegraphics[width=80mm]{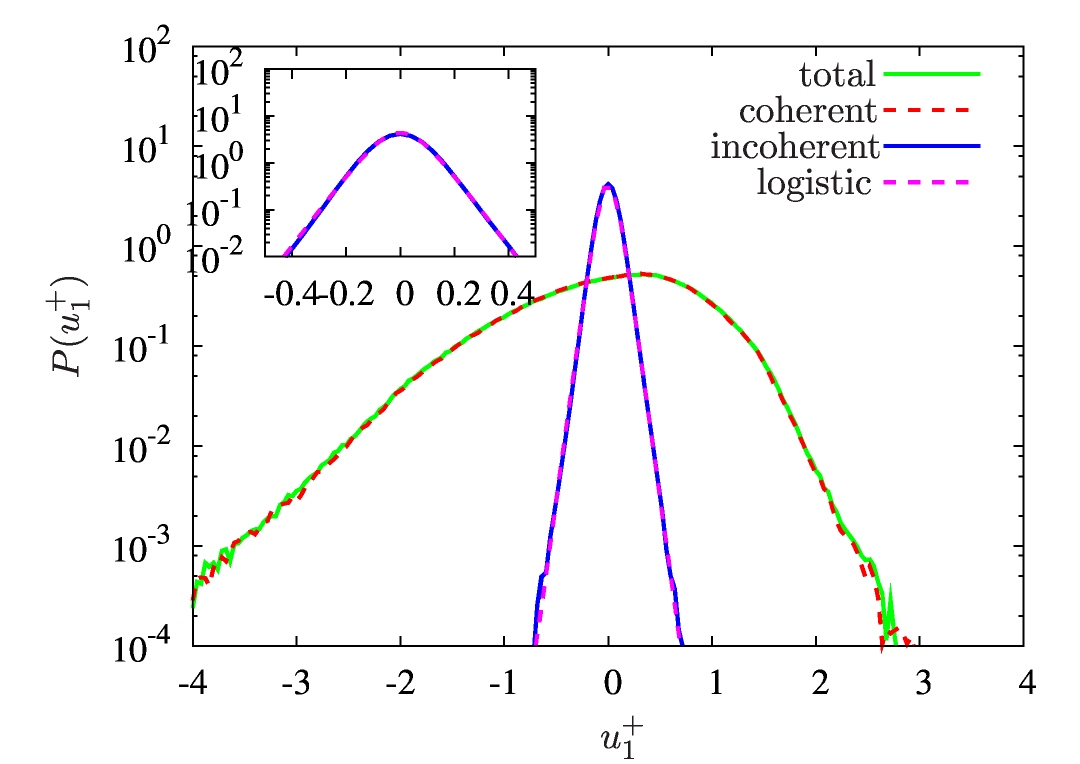}
\caption{PDFs of $u_1^+$; $x_2^{+}=378.8$ around the center of the channel with CVE using Chebyshev wavelets. }
\label{Cheb_PDF}
\end{center}
\end{figure}

Now we extract coherent vorticity out of the turbulent channel flow at $Re_\tau=395$, 
using the previously described Chebyshev wavelets. 
For the threshold value $T$ we use the coefficient $\alpha=0.10$.
We find that the coherent flow, 
reconstructed from only $ 4.8\% $ of the $256^2\times 512$ wavelet coefficients, 
i.e., $6.4\%$ of the original $256^2 \times 192$ grid points,
retains almost all of the total energy and enstrophy, 
i.e., $99.9 \%$ of the total energy and $99.0 \%$ of the total enstrophy.
In contrast, the incoherent flow represented by the remaining majority of the wavelet coefficients 
has little energy and enstrophy, namely $10^{-2} \, \%$ of the total energy and $1.3 \%$ of the total enstrophy. 

Inspecting Fig. \ref{Cheb_PDF} confirms that the PDFs for the total and coherent velocity fluctuations perfectly superimpose, 
indicating that high order statistics are well preserved by the coherent flow.
In contrast the PDFs of incoherent velocity fluctuations have strongly reduced variances and 
are not skewed, in contrast to what is found for the total and coherent fluctuations.
Coherent and incoherent flows exhibit very similar properties as in Sec. \ref{sec3}, 
where we used CDJV wavelets instead of the Chebyshev wavelets (figure with flow visualizations is omitted).
Thus, Chebyshev wavelets can be more efficient for CVE than CDJV wavelets if the flow data have a large number of grid points, as no interpolation onto a fine equidistant grid is required.


\end{document}